\documentclass[useAMS,usenatbib]{mn2e}
\usepackage{graphicx}

\title[Study of the W40 cloud]{
Molecular line and continuum study of the W40 cloud}

\author[L. Pirogov et al.]
{L. Pirogov$^1$
\thanks{E-mail: pirogov@appl.sci-nnov.ru},
D.K. Ojha$^2$, M. Thomasson$^3$, Y.-F. Wu$^4$, I. Zinchenko$^1$\\
$^1$Institute of Applied Physics RAS, Ulyanova 46, Nizhny Novgorod 603950, Russia\\
$^2$Department of Astronomy and Astrophysics, Tata Institute of Fundamental Research, Homi Bhabha Road, Mumbai 400005, India \\
$^3$Chalmers University of Technology, Department of Radio and Space Science, Onsala Space Observatory, 43992 Onsala, Sweden \\
$^4$Department of Astronomy, School of Physics, Peking University, Beijing, 100871, China
}

\begin{document}

\date{Accepted 2013 September 22. Received 2013 September 18; in original form 2013 July 22}

\pagerange{\pageref{firstpage}--\pageref{lastpage}} \pubyear{2013}

\maketitle

\label{firstpage}

\begin{abstract}
The dense cloud associated with W40, one of the nearby H~II regions,
has been studied in millimeter-wave molecular lines
and in 1.2~mm continuum.
Besides, 1280~MHz and 610~MHz interferometric observations
have been done.
The cloud has complex morphological and kinematical structure,
including a clumpy dust ring and an extended dense core.
The ring is probably formed by the ``collect and collapse'' process
due to the expansion of neighboring H~II region.
Nine dust clumps in the ring have been deconvolved.
Their sizes, masses and peak hydrogen column densities
are: $\sim 0.02-0.11$~pc,
$\sim 0.4-8.1~M_{\odot}$ and $\sim (2.5-11)\times 10^{22}$~cm$^{-2}$,
respectively.
Molecular lines are observed at two different velocities
and have different spatial distributions
implying strong chemical differentiation over the region.
The CS abundance is enhanced towards the eastern dust clump~2,
while the NH$_3$, N$_2$H$^+$, and H$^{13}$CO$^+$ abundances
are enhanced towards the western clumps.
HCN and HCO$^+$ do not
correlate with the dust probably tracing the surrounding gas.
Number densities derived towards selected positions
are: $\sim (0.3-3.2)\times 10^6$~cm$^{-3}$.
Two western clumps
have kinetic temperatures 21~K and 16~K and are close to virial equilibrium.
The eastern clumps 2 and 3 are more massive,
have higher extent of turbulence and are probably more evolved
than the western ones.
They show asymmetric CS(2--1) line profiles
due to infalling motions
which is confirmed by model calculations.
An interaction between ionized and neutral material is taking place
in the vicinity of the eastern branch of the ring
and probably trigger star formation.

\end{abstract}


\newpage

\section{Introduction}

The process of high-mass star formation attracts much attention nowadays.
There is still no general theory of high-mass star formation
and evolution (e.g. Zinneker \& Yorke 2007)
as in the case of low-mass star formation.
The regions where massive stars are forming are more rare,
more distant and evolve more quickly than low-mass star-forming regions.
During their evolution massive stars affect the surrounding
parent cloud due to stellar winds, massive outflows,
strong UV radiation, and expansion of H~II regions.
These strong impacts affect the physical conditions
and chemical composition of the parent cloud and may cause contraction
and trigger the formation of new generation of stars.
Therefore, the detailed studies of high-mass star-forming (HMSF)
regions are important.

One of the nearby regions of high-mass star formation, W40
also known as Sharpless 64 (Westerhout 1958, Sharpless 1959),
contains a large blister-type H~II region ($\sim 1$~pc in diameter)
which lies at the edge of extended molecular core
(TGU 279-P7 according to Dobashi et al. 2005)
in the Aquila Rift complex at the Galactic latitude $b\sim 3.5\degr$.
The region has been extensively studied at different wavelengths.
There is an embedded cluster of young stars in the center of the region.
A review of the existing data and the properties of the H~II region,
inner sources and molecular cloud is given in Rodney \& Reipurth (2008).
More recent observational data of the region are obtained
in the radio (Rodr\'iguez et al. 2010), NIR (Shuping et al. 2012)
and X-ray (Kuhn et al. 2010) ranges.
The Aquila Rift complex including the W40 region has been observed
in millimeter continuum by Maury et al. (2011).
A survey of W40 on large scales on the base
of original and archival multiwavelength data has been done recently
by Mallick et al. (2013).

The distance to W40 region is not exactly constrained.
Atomic and molecular line observations
enabled to deduce kinematical distance
of 600$\pm$300~pc (Radhakrishnan et al. 1972, see also
discussions in Vall\'ee 1987, Rodney \& Reipurth 2008).
The ``far'' value of the two-fold ambiguity
in the kinematical distance calculations is rejected
on the basis of optical identification
and high distance from the Galactic plane (Reifenstein et al. 1970).
Bontemps et al. (2010) associated W40 with the Serpens star-forming region
in the Aquila Rift complex and suggested a distance of 260~pc.
Kuhn et al. (2010) used the XLF fitting method
and derived a best fit distance value of 600~pc.
Recently, Shuping et al. (2012) from the analysis of the NIR data
have estimated the distance to three stars of the W40 cluster
to lie between 455~pc and 536~pc.
They ruled out distances less than 340~pc and higher than 686~pc.
We adopt a distance of 500~pc throughout our paper.
This puts W40 among most closely located HMSF regions.

The associated W40 molecular cloud is located $\sim 2'$
to the west of the H~II region.
It was observed in molecular and atomic lines by different authors.
Early CO(1--0) observations
by Zeilik \& Lada (1978) revealed a region of extended emission
at $\sim 4.5$~km~s$^{-1}$ with two emission peaks
located in the north-south direction.
They argued that the velocity difference
between CO and hydrogen recombination lines ($\sim 4$~km~s$^{-1}$)
is consistent with the blister model in which H~II region
lies at the front edge of the cloud closer to the observer.
Based on the carbon recombination line
observations at $\sim 6.5$~km~s$^{-1}$ Vall\'ee (1987)
proposed the model according to which
the western portion of the H~II region is obscured
by molecular cloud.
Crutcher (1977) observed the OH absorption lines at $\sim 7$~km~s$^{-1}$
towards the W40 H~II region.
The molecular cloud was mapped
in CO, $^{13}$CO, HCO$^+$, HCN
and H$\alpha$ by Crutcher \& Chu (1982).
They revealed that the CO and $^{13}$CO line profiles possess
two emission features at $\sim 5$ and $\sim 8$~km~s$^{-1}$.
These two velocity components have different spatial distributions.
The HCN and HCO$^+$ maps revealed the existence of a high-density core.
The HCO$^+$(1--0) line widths ($\sim 1.4$~km~s$^{-1}$,
Pirogov et al. 1995)
imply moderate degree of turbulence of the dense gas
compared with typical HMSF regions (Pirogov et al. 2003).
The CO isotopic line observations by Zhu et al. (2006)
allowed to determine the mass of the W40 core
$\sim 200-300$~$M_{\odot}$ which is close to model estimates
from Vall\'ee et al. (1992) and typical for HMSF cores.
Weak CO outflow is detected in the center of the core by Zhu et al. (2006).
Yet, the driving source of the outflow remained undiscovered.
Mallick et al. (2013) derived lower limit on the mass contained
within the area of the 3$'$ radius around the central sources
of 126~$M_{\odot}$.

Our goal is to study
the structure of W40 cloud in the vicinity of the H~II region
where interaction between ionized and
dense neutral material is taking place.
It is important to derive gas and dust physical parameters
as well as chemical composition of the gas where formation
of the next generation of stars is probably taking place.

\section{Observations}
\label{observations}

We performed molecular multiline and dust continuum observations
of the W40 cloud at millimeter wavelengths using IRAM-30,
Onsala-20, Effelsberg-100 telescopes.
Besides, the region was observed at 1280~MHz and 610~MHz (UHF band)
with Giant Metrewave Radio Telescope (GMRT) array.
Our maps are centered at the position
R.A.(J2000)=18$^{\rmn{h}}$~31$^{\rmn{m}}$~15.75$^{\rmn{s}}$,
Dec.(J2000)=--02$\degr$~06$'$~49.3$''$ which is close to the
CO and $^{13}$CO peak position from Zhu et al. (2006).
The
parameters of single-dish molecular line observations
are given in Table~\ref{tel_param}.

\begin{table*}
\centering
\caption[]{The parameters of single-dish molecular line observations.}
\begin{tabular}{lccccc}\hline\noalign{\smallskip}
Telescope     &
Frequency     &
$\Delta \Theta$  &
$\eta_{\rm mb} $ &
$T_{\rm sys}$       &
$\delta V$         \\
 & (GHz)      & ($\arcsec$) &           & (K)       & (km s$^{-1}$)\\
\noalign{\smallskip}\hline\noalign{\smallskip}
IRAM-30 & 85--99  & 29--26    & $\sim 0.8$         & 110--140  & 0.24--0.27 \\
        & 245     & 10        & $\sim 0.57$        & 400--600  & 0.38 \\
OSO-20  & 86--96  & $\sim 40-44$  & $\sim 0.38-0.45$   & 200--700  & 0.078-0.087\\
Effelsberg-100 & 23.7 & 40        & 0.53               &  70--110  & 0.039 \\
\noalign{\smallskip}\hline\noalign{\smallskip}
\end{tabular}
\label{tel_param}
\end{table*}

\subsection{IRAM-30 observations}

Mapping of W40 cloud in the CS(2--1), CS(5--4), N$_2$H$^+$(1--0)
and CH$_3$CCH(5--4) molecular lines has been performed in August 2010
at the IRAM-30 telescope with Eight Mixer Receiver (EMIR)
(the E090/E230 bands combination).
The Vespa autocorrelator with 0.08/0.32 MHz spectral resolutions
for E090/E230 bands, respectively, was used as a backend.
In addition, the Wilma autocorrelator with 2~MHz spectral resolution
was used.
The maps were taken in the On-The-Fly (OTF) mode and covered
a $6'\times 4'$ region around the central position
(the region observed in N$_2$H$^+$(1--0) is $6'\times 2'$).
The reference position was taken at (-1500$''$, 0$''$).
In the further analysis the OTF maps have been gridded
with different steps ranging from $4''$ to 20$''$.
Pointing and focus
were checked on nearby strong continuum sources and Mars.

The 1.2 mm continuum observations have been done in autumn 2010
during pool session with MAMBO2 bolometer array in the On-The-Fly mode.
The antenna HPBW for these observations is about 11$''$.
The data have been reduced by the $mopsic$ package
both with
skynoise filtering.
The mean r.m.s. noise level calculated from the analysis
of beam sampled regions without noticeable emission
is $\sim 10$~mJy~beam$^{-1}$.

\subsection{OSO-20 observations}

The observations with the 20m telescope
of the Onsala Space Observatory
have been done in February 2011.
The HCN(1--0) and HCO$^+$(1--0) maps have been obtained.
Distinct positions have been observed in the H$^{13}$CN(1--0),
H$^{13}$CO$^+$(1--0), C$^{34}$S(2--1) and CH$_3$OH(2--1) lines.
We used superconductor-insulator-superconductor (SIS) receiver
in the frequency-switching mode with
autocorrelation spectrometer as a backend with 20~MHz/25~kHz
bandwidth and spectral resolution, respectively.
The HCN and HCO$^+$ maps cover approximately $4'\times 3'$
around the central position skipping the eastern-northern and
the southern-western corners of this rectangular region
where lines are relatively weak.
Mapping has been done with $20''$ grid spacing (Nyquist sampled maps).
Pointing was checked periodically by observations of R~Cas SiO maser
and usually is better than $5''$.
System temperatures were $\sim 300-500$~K for HCN(1--0),
$\sim 400-700$~K for HCO$^+$(1--0) and $\sim 200-300$~K
for other molecular line observations.

\subsection{Effelsberg-100 observations}

The NH$_3$ (1,1) and (2,2) observations with Effelsberg-100
telescope have been done in April 2012.
Both transitions were observed simultaneously using receiver with
cooled HEMT amplifier.
The XFFTS spectrometer with total bandwidth of 100~MHz and
3.05~kHz of spectral resolution has been used.
Mapping has been done in the raster mode with grid spacing of 20$''$
(Nyquist sampling) in the frequency-switching mode.
The region of observations was divided into two neighboring subregions
with sizes: $140''\times 160''$ and $100''\times 280''$ to the east and
to the west of the central position, respectively.
Pointing and focus have been checked periodically on both strong
and nearby continuum sources. Pointing accuracy is $\la 8''$.
In order to convert the observed intensities into main beam
temperatures we took into account the opacity and gain-elevation corrections
(A.Kraus, private communication).
The primary calibration source NGC~7027 was used.
The flux density was taken from Ott et al. (1994).

\subsection{GMRT observations}

The radio continuum observations at 1280~MHz and at 610~MHz
were performed with the GMRT array on November 15 and 18, 2011.
The GMRT consisted of 30 antennae each having 45-m in diameter
which are located in an ``Y-shaped'' hybrid configuration.
A central region of about 1~square km consists
of 12 randomly distributed antennae.
The rest are located along three radial arms up to $\sim 14$~km long
(more technical details can be found in Swarup et al. 1991).
The primary GMRT beams at 1280~MHz and 610~MHz are 24$'$ and 43$'$,
respectively.
At both frequencies 3C286 and 3C48 were used as flux calibrators,
1822-096 was used as phase calibrator.
The data analysis was done with the AIPS package including
flagging corrupted data, calibration, Fourier inversion and cleaning.
The resulting images have synthesized beams at 1280 MHz and 610 MHz
of $2.60''\times 2.25''$ and $5.24''\times 4.79''$, respectively.
The resulted r.m.s are 0.16~mJy~beam$^{-1}$ and 0.75~mJy~beam$^{-1}$
at 1280~MHz and at 610~MHz, respectively.

The comparison with the coordinates of the 3.6~cm VLA sources
observed towards the W40 central region by Rodr\'iguez et al. (2010)
reveals systematic shifts of $\sim 1.5''$ and $\sim 4.5''$
between the VLA and the associated sources at 1280~MHz and 610~MHz,
respectively.
The reasons for these shifts are unclear but probably are
related to phase calibration at GMRT.
We applied corresponding positional corrections
to the GMRT maps to compensate the shifts.

\section{Results and data analysis}
\label{obs_results}

\subsection{1.2 mm continuum data}
\label{mm_cont_data}

\begin{figure*}
    \includegraphics[width=0.6\textwidth,angle=-90]{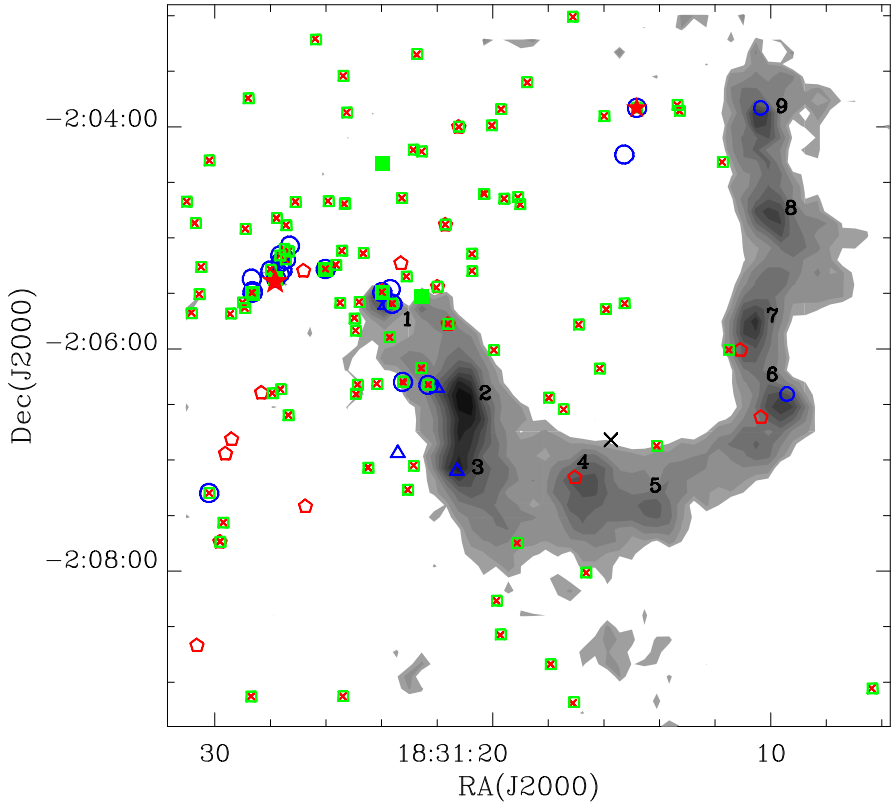}
\caption{\small
The W40 molecular cloud in 1.2~mm dust continuum (grey scale).
Intensity contours range from 30~mJy~beam$^{-1}$ (3$\sigma$)
to 210~mJy~beam$^{-1}$ with the 10~mJy~beam$^{-1}$ step
plus the 20~mJy~beam$^{-1}$ contour in addition.
The peak flux is 219.5~mJy~beam$^{-1}$.
The dust clumps for which we derived physical parameters are marked by numbers.
Compact VLA 3.6~cm sources from Rodr\'iguez et al. (2010)
(big blue circles), NIR sources from Shuping et al. (2012)
(filled green squares) as well as 2MASS sources (empty green squares)
coincided with X-ray sources (Kuhn et al. 2010) (red crosses)
are shown.
The Class~0 and Class~I sources from Maury et al. (2011)
are shown as small blue circles and triangles, respectively.
The Class~I sources from Mallick et al. (2013) are shown
as red pentagons.
The peak $^{13}$CO position (Zhu et al. 2006) is indicated
as a black cross.
The main driving source of the H~II region
(IRS~1A~South) and the distinct IRS~5 source (Shuping et al. 2012)
are marked by bigger and smaller red stars, respectively.
Color figures are available in the online version.
}
\label{dust_sources}
\end{figure*}

The dust continuum emission map of the W40 region at 1.2~mm
is shown in Fig.~\ref{dust_sources}.
The map reveals a ring-like structure consisting of dust clumps.
This structure has been previously detected by Maury et al. (2011)
in their 1.2~mm continuum study of the Aquila rift complex.
The 3.6~cm VLA sources (Rodr\'iguez et al. 2010)
and NIR sources (Shuping et al. 2012) are also shown
in Fig.~\ref{dust_sources}.
Most of the NIR sources coincide with those detected earlier
by Smith et al. (1985) being the members of the central cluster.
Using the NIR spectroscopic data Shuping et al. (2012) derived
spectral types of the sources and considered the IRS~1A~South source (O9.5 type)
to be the dominant source of LyC photons needed
to power of the H~II region (as opposed to IRS~2A originally
suggested by Smith et al. 1985).
One of the NIR sources, IRS~5, of the B1V spectral type
(Shuping et al. 2012, Mallick et al. 2013) probably
does not belong to the cluster.
The IRS~5 source has a counterpart on the VLA maps (Rodr\'iguez et al. 2010).
Multiple NIR (Rodr\'iguez et al. 2010), 2MASS, MSX mid-IR sources,
FIR sources (detected by $Herschel$, Bontemps et al. 2010)
and Chandra X-ray sources (Kuhn et al. 2010) are also detected
in the region.
The 2MASS and X-ray sources with position differences  $\le 1''$
that could be tracers of young stellar objects embedded in the cloud
are also shown in Fig.~\ref{dust_sources}.
Most of these sources are located outside the ring in the east-north
part of the map but some of them can be associated
with the eastern branch of the ring.

We deconvolved the dust map into nine individual clumps using our
2D Gaussian fitting program (Pirogov et al. 2003).
The clumps are marked by numbers (Fig.~\ref{dust_sources}).
Clumps~1 and 9 have dimensions lower than antenna HPBW.
Parameters of individual clumps are given in Table~\ref{dust_clumps}.
It includes relative offsets of clump centers
with respect to the central position,
aspect ratios,
deconvolved angular and linear sizes estimated as geometric mean
of the extents of elliptical Gaussians (Cols. 2--6).
For the clump~9
the results of 2D circular Gaussian fit are given.
Total fluxes of the clumps are given in Col.~7.
For the clumps 2 and 3 which are closely located intensity enhancements
over the local plateau approximately elongated in the north-south direction
in the eastern branch of the ring the total flux is given.
Most strong 1.2~mm continuum emission comes from the eastern clumps
which are closer to the cluster sources.
The IRS~2A and VLA~5--7 sources are associated with
the clump~1.
The VLA~3 source is located near the clump~2.
The other clumps have no associated IR or radio sources.

\begin{table*}
\centering
\caption[]{Parameters of 1.2 mm continuum clumps}
\small
\begin{tabular}{lrrrrrrr}
\noalign{\hrule}\noalign{\smallskip}
Object   & $\Delta\alpha$  & $\Delta\delta$ & Aspect  &$\Delta \Theta$ & $d$ & $F_{\rm total}$ & Associated \\
         &  ($\arcsec$)    &  ($\arcsec$)   & ratio  &  ($\arcsec$)   &(pc) & (mJy)            & sources \\
\noalign{\smallskip}\hline\noalign{\smallskip}

Clump 1     & 124.2(0.3) &   72.7(0.2)   &1.8(0.5)  &  9.6(1.2) & 0.023(0.003) & 188(7)       & Class I$^b$ \\
Clump 2     &  79.3(0.2) &   19.8(0.5)   &2.0(0.1)  & 26.3(0.8) & 0.064(0.002) & 2910(21)$^a$ & Class I$^b$ \\
Clump 3     &  80.4(0.3) & --12.6(0.7)   &1.4(0.2)  & 24.9(1.5) & 0.060(0.004) &              & Class I$^b$ \\
Clump 4     &  12.1(0.3) & --29.7(0.6)   &1.2(0.1)  & 45.1(2.0) & 0.109(0.005) & 1310(15)     & Class I$^c$\\
Clump 5     &--20.1(0.5) & --36.4(0.4)   &1.3(0.2)  & 30.7(1.8) & 0.074(0.004) & 519(10) \\
Clump 6     &--90.0(0.3) &   18.5(0.4)   &1.3(0.2)  & 24.2(1.8) & 0.059(0.003) & 737(11) & Class 0$^b$\\
            &            &               &          &           &              &         & Class I (?)$^c$\\
Clump 7     &--77.0(0.2) &   62.5(0.4)   &2.0(0.3)  & 22.7(1.4) & 0.055(0.003) & 527(8)  & Class I (?)$^c$ \\
Clump 8     &--88.0(0.3) &  121.6(0.4)   &1.9(0.2)  & 27.9(1.3) & 0.068(0.003) & 792(11) &  \\
Clump 9     &--80.8(0.4) &  173.2(0.4)   &          &  5.8(2.1) & 0.014(0.005) & 570(9)  & Class 0$^b$\\
\noalign{\smallskip}\hline\noalign{\smallskip}
\end{tabular}

\small
\flushleft{
$^a$ -- total flux for the clumps 2 and 3 \\
$^b$ -- Maury et al. 2011 \\
$^c$ -- Mallick et al. 2013
}

\label{dust_clumps}
\end{table*}

Maury et al. (2011) have found in total 36 mm-continuum sources
associated with the W40 region, 11 of which are located within
the region observed by us.
Three more sources from their survey
are located outside our field of view to the north of IRS~5
but probably belong to the ring.
Using the mm-continuum and $Herschel$ data they derived
several physical parameters of the sources.
In the further analysis we used their dust temperature estimates
for the clumps which are associated with their continuum sources.
Several Class~I sources from Mallick et al. (2013)
are also located within the observed region.
In the last column of Table~\ref{dust_clumps} associations
with the Class~I and Class~0 sources
found by Maury et al. (2011) and Mallick et al. (2013) are given.
The western dust clumps~6 and 9 are associated
with Class~0 sources
while the eastern clumps 2 and 3 can be associated
with Class~I sources (Maury et al. 2011).
The Class~I sources from Mallick et al. (2013) are located
near the dust clumps 6 and 7 but are shifted
from their centers.
The association of them with clumps is questionable.

\subsection{Molecular line data}
\label{mol_data}

The molecular line maps of the W40 cloud show different
morphologies implying rather complicated structure of the region.
Among the observed molecular lines no one appears to trace
all dust clumps of the ring.
Some species show correlations with distinct dust clumps
but some of them do not correlate with dust continuum.
Molecular emission is also found in the area
with no prominent dust emission implying that dense gas and dust
do not follow each other.

\begin{figure*}
\begin{minipage}[b]{0.49\textwidth}
  \includegraphics[width=\textwidth,angle=-90]{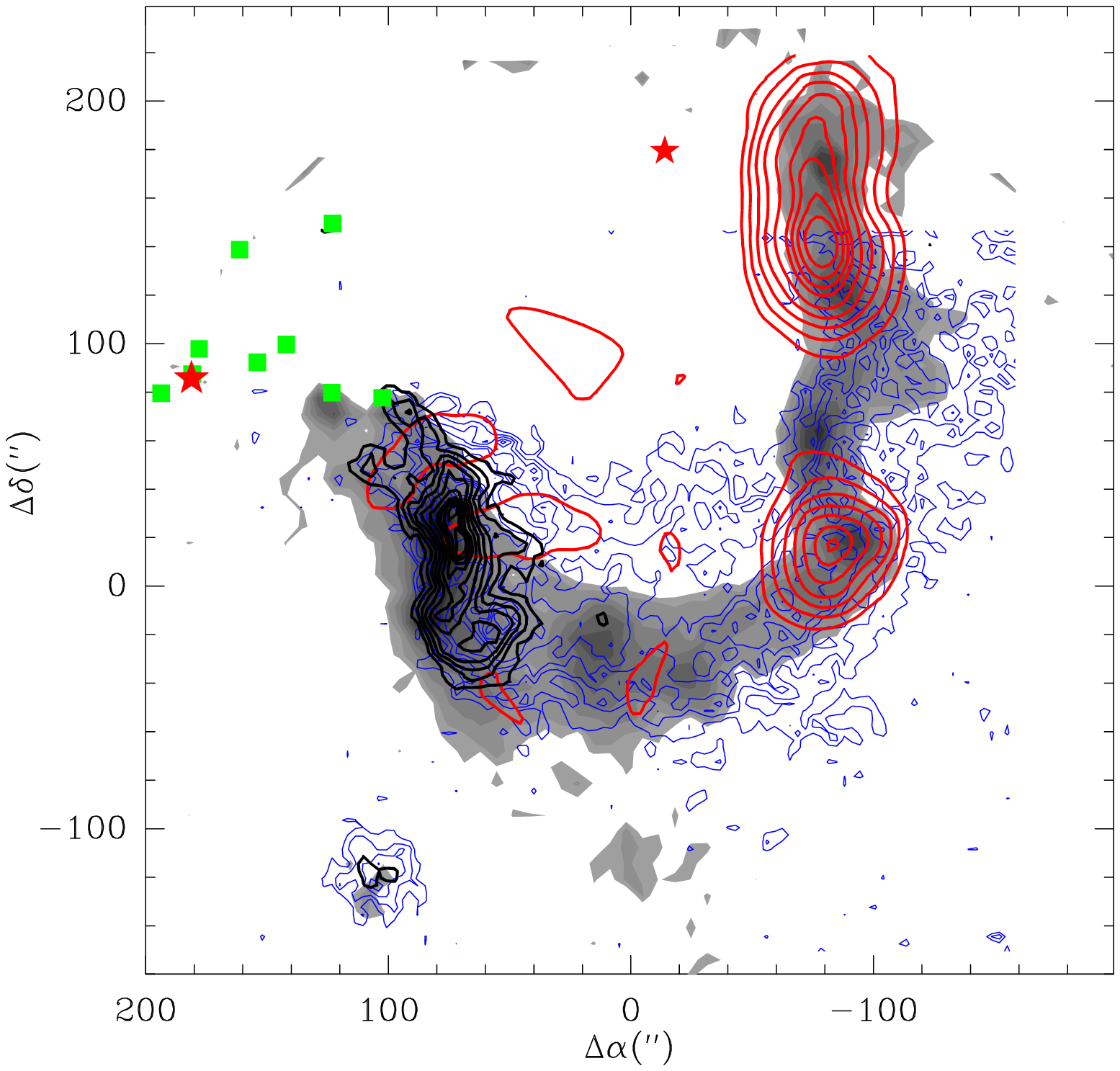}
\end{minipage}
\hspace{1mm}
\begin{minipage}[b]{0.49\textwidth}
  \includegraphics[width=\textwidth,angle=-90]{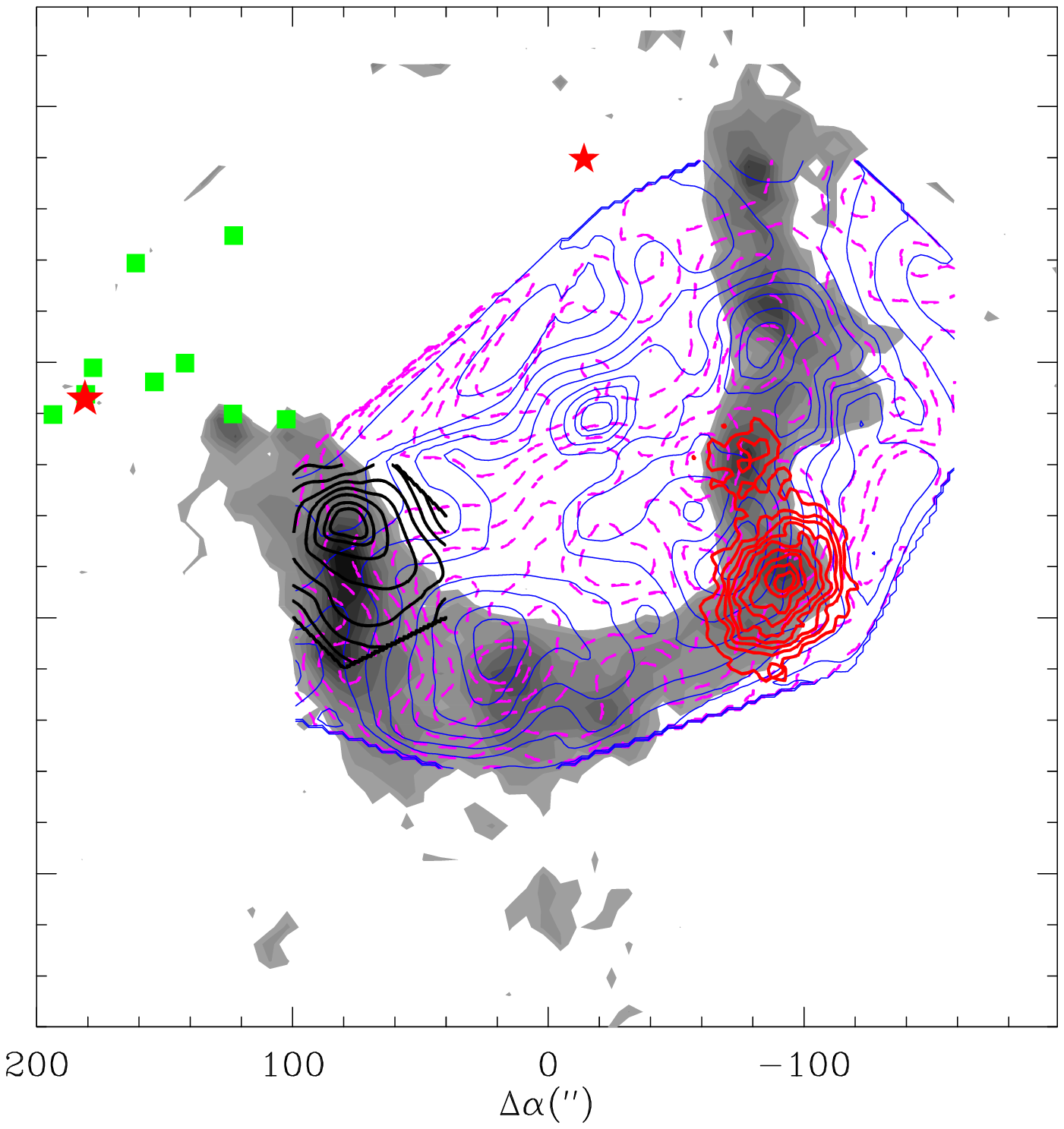}
\end{minipage}

\caption{\small
Molecular line integrated intensity maps
overlayed over 1.2~mm dust continuum emission (greyscale).
Left panel: CS(2--1) (blue contours), NH$_3$(1,1) (red contours),
CS(5--4) (black contours).
Right panel: HCN(1--0) (crimson dashed contours), HCO$^+$(1--0) (blue contours),
N$_2$H$^+$(1--0) (red contours) and C$^{34}$S(2--1) (black contours).
Intensity contours range from 20\% to 90\% with the step of 10\% of the peak
values plus the contour of 95\%.
The peak integrated intensities (in K~km~s$^{-1}$) are:
14.5 for CS(2--1), 36.5 for CS(5--4), 16.1 for NH$_3$(1,1), 15.6 for HCN(1--0),
10.9 for HCO$^+$(1--0), 22.7 for N$_2$H$^+$(1--0) and 2.2 for C$^{34}$S(2--1).
Coordinates are the offsets with respect to the position
of the $^{13}$CO peak (Zhu et al. 2006).
The NIR sources are marked by squares, the main driving source
of the H~II region (IRS~1A~South) is marked by
bigger red star, the IRS~5 source is marked by smaller red star.
}
\label{line_maps}
\end{figure*}

\subsubsection{Molecular line maps}

Molecular integrated intensity maps
overlayed over the dust continuum image are shown in Fig.~\ref{line_maps}.
Only the eastern part of the region has been mapped in C$^{34}$S(2--1),
in the other parts the C$^{34}$S(2--1) observations
have been done towards distinct positions.
The CS lines are strong towards
the eastern branch of the ring while the N$_2$H$^+$(1--0)
and the NH$_3$(1,1) lines are strong towards the western branch.
The CS(5--4) map correlates with the eastern branch.
The CS(2--1) map follows the dust ring in shape but is more extended.
The HCN(1--0) and HCO$^+$(1--0) maps
cover an extended part of the region ($\sim 3'\times 2.5'$)
and show practically no correlation with dust.
The H$^{13}$CO$^+$(1--0), H$^{13}$CN(1--0) and CH$_3$OH(2--1) lines
are observed towards distinct positions.
The CH$_3$CCH(5--4) emission has not been detected
in the region at 3$\sigma$ level of 0.06~K.

The N$_2$H$^+$(1--0), NH$_3$(1,1), CS(5--4) and C$^{34}$S(2--1) maps
are deconvolved into clumps in the similar way as the dust clumps
(Section~\ref{mm_cont_data}).
We have also detected the CS clump which lies to the south
of the dust ring and is not associated with it (Fig.~\ref{line_maps}).
Weak dust emission can be associated with this clump.
No attempts are made to deconvolve the CS(2--1) clumps
towards the eastern branch of the ring where
the line profiles are multicomponent.
Towards the other parts of the cloud the CS(2--1) lines
do not trace distinct clumps.
The CS(5--4) clumps~1 and 2 are deconvolved
from the intensity maps integrated in different velocity ranges.
The derived parameters of molecular clumps are given in Table~\ref{mol_clumps}.
In the last column the associations with dust clumps are given.
The centers and sizes of the N$_2$H$^+$ clump, the ammonia clump~1
and the dust clump~6 are close to each other.
The ammonia clump~2 extends over both dust clumps~8 and 9
with the center located between these dust clumps.
Probably the spatial resolution of the ammonia observations
was not sufficient to resolve these closely located clumps.
The area containing dust clumps~8 and 9
has not been observed in N$_2$H$^+$(1--0).
The CS(5--4) clump~1 and the C$^{34}$S clump are associated
with the dust clump~2.

\begin{table*}
\centering
\caption[]{Parameters of molecular clumps}
\small
\begin{tabular}{lrrrrrr}
\noalign{\hrule}\noalign{\smallskip}
Object   & $\Delta\alpha$  & $\Delta\delta$ & Aspect  &$\Delta \Theta$ & $d$ & Associations \\
         &  ($\arcsec$)    &  ($\arcsec$)   & ratio  &  ($\arcsec$)   &(pc)  & with dust clumps \\
\noalign{\smallskip}\hline\noalign{\smallskip}

N$_2$H$^+$ clump  & --90.1(0.2) &   12.8(0.2)   &1.8(0.1)  & 29.7(0.5) & 0.072(0.001) & clump 6 \\
Ammonia clump 1  & --82.4(1.1) &  17.3(1.3)   &2.2(0.9)  & 27.6(5.4)    & 0.067(0.013) & clump 6 \\
Ammonia clump 2  & --77.9(0.8) & 157.0(1.8)   &4.7(1.3)  & 42.7(6.1)    & 0.104(0.015) & clumps 8 \& 9   \\
CS(5--4) clump 1$^a$   & 72.6(0.2) &   18.5(0.7)   &3.0(0.2)  & 30.9(0.8)  &  0.075(0.002) &  clump 2 \\
CS(5--4) clump 2$^b$   & 54.0(0.4) & --15.4(0.3)   &1.7(0.2)  & 15.0(0.8)  & 0.036(0.002)  &  clump 3 (?) \\
C$^{34}$S clump   & 71.3(2.8) &  26.2(2.7)   &1.5(0.5)  & 49.4(8.0)    & 0.12(0.02) & clump 2 \\
Southern CS clump & 105.5(0.4) & --118.4(0.4) &1.3(0.2) & 19.6(1.3) & 0.048(0.003) &  \\
\noalign{\smallskip}\hline\noalign{\smallskip}
\end{tabular}

\small
\flushleft{
$^a$ -- integrated in the [6...9]~km~s$^{-1}$ velocity range \\
$^b$ -- integrated in the [3...6]~km~s$^{-1}$ velocity range
}

\label{mol_clumps}
\end{table*}

\subsubsection{Analysis of molecular spectra}
\label{mol_spectra}

After removing baselines from the data (low-order polynomials)
line parameters have been obtained from Gaussian fitting.
The Table~\ref{mol_peaks} includes integrated intensities
and parameters of Gaussian fits (main beam temperature,
velocity, and line width) with corresponding 1~$\sigma$ errors
towards selected positions.
The positions are close
to the CS, N$_2$H$^+$ or NH$_3$ (1,1) local integrated intensity peaks
and corresponding clump centers.
We also include the central position
and positions to the north and to the south with respect
to it (0$''$,$\pm 40''$).
For the purpose of comparison with the Onsala and Effelsberg data
the IRAM data have been averaged over areas with 20$''$ radius
around given positions.
The CS spectra towards the position of the ``southern CS'' clump
observed only at IRAM
are averaged for the area with the radius of 10$''$.
No fitting has been done for the CS(2--1) lines
towards the $(80'',40'')$ and $(80'',0'')$ positions
as the profiles are multicomponent.

The N$_2$H$^+$(1--0) and the NH$_3$ spectra observed
towards the western part of the region
show multiple and partly overlapping hyperfine components.
These spectra are fitted with the function consisting
of 7, 18 and 24 components
for N$_2$H$^+$(1--0), NH$_3$ (1,1) and NH$_3$(2,2), respectively.
Excitation temperatures are assumed to be the same
within each group of overlapping components but can differ
for different groups
(three and five for N$_2$H$^+$ and NH$_3$, respectively).
The method was used previously for fitting the N$_2$H$^+$(1--0) spectra
(Pirogov et al. 2003).
The fitting parameters are: the number of excitation temperatures
equal to the number of groups of hyperfine components,
velocity of the central component, line width and total optical depth.
For the $(-100'',20'')$ and $(-80'',140'')$ positions
(Table~\ref{mol_peaks}) the N$_2$H$^+$(1--0) and NH$_3$(1,1)
excitation temperatures of the main (central) group
of hyperfine components
are given instead of main beam temperatures.

The HCN(1--0) spectra consist of three hyperfine components
and have been fitted by three Gaussians of the same width.
The $F=1-1$ HCN(1--0) hyperfine component (the red one)
in the central and western parts of the observed region
is systematically shifted with respect to the expected position.
The profiles of individual hyperfine components often deviate
from Gaussian.
Probably, different HCN(1--0) hyperfine components trace gas
with slightly different velocities.
As the signal-to-noise ratios are not high, fitting of the HCN(1--0) spectra
by more than one closely located Gaussian triplets is often ambiguous.
Therefore, the results of fitting by single triplet
are given in Table~\ref{mol_peaks}.
Main beam temperature corresponds to the central hyperfine component $F=2-1$.
We don't use these results in the further analysis.

\smallskipamount 0.3pt

\begin{table*}
\centering
\caption[]{Molecular line parameters towards selected positions}
\small
\begin{tabular}{llrrrrr}
\noalign{\hrule}\noalign{\smallskip}
Offsets &Line    &$I$            & $T_{\rm MB}$  & $V_{\rm LSR}$  & $\Delta V$     & $\tau$ \\
$('','')$&        &(K km s$^{-1}$)&  (K)          &  (km s$^{-1}$) &  (km s$^{-1}$) \\
\noalign{\smallskip}\hline\noalign{\smallskip}
(120, 60) &CS(2--1)        & 0.5(0.1) & 0.26(0.03) & 5.7(0.1)   & 1.8(0.2) \\
             &                & 1.9(0.1) & 0.67(0.04) & 8.92(0.03) & 1.2(0.1) \\
             &CS(5--4)        & 3.0(0.1) & 1.4(0.1)   & 8.30(0.04) & 2.0(0.1) \\
             &N$_2$H$^+$(1--0)&          & $<0.15 $ \\
             &NH$_3$(1,1)     &          & $<1.2$ \\
\noalign{\smallskip}\hline
(80, 40)  &CS(2--1)        & 8.1(0.1) \\
             &CS(5--4)        & 15.1(0.2)  & 5.5(0.1)   & 7.37(0.02) & 2.62(0.04) \\
             &C$^{34}$S(2--1) & 1.9(0.1)   & 0.74(0.03) & 7.07(0.05) & 2.59(0.12) \\
             &N$_2$H$^+$(1--0)& 0.7(0.1)   & 0.19(0.03) & 6.97(0.06) & 0.7(0.1)   \\
             &NH$_3$(1,1)     & 1.5(0.2)   & 0.39(0.04) & 6.4(0.1)   & 1.8(0.2)   \\
             &HCN(1--0)       & 7.0(0.5)   & 1.9(0.1)  & 5.0(0.1) & 1.8(0.1) \\
             &HCO$^+$(1--0)   & 2.8(0.4)   & 2.0(0.2)  & 4.5(0.1) & 1.6(0.2) \\
             &H$^{13}$CN(1--0) & 0.9(0.1)  & 0.5(0.1)  & 7.1(0.1) & 1.4(0.3) \\
\noalign{\smallskip}\hline
(80, 0)   &CS(2--1)        & 5.5(0.1) \\
             &CS(5--4)        & 16.2(0.1)  & 8.0(0.1)  & 7.83(0.01) & 1.71(0.03) \\
             &C$^{34}$S(2--1) & 1.4(0.1)   & 0.56(0.03) & 7.68(0.05) & 2.0(0.1) \\
             &N$_2$H$^+$(1--0)& 1.3(0.1)   & 0.26(0.03) & 8.03(0.04) & 1.0(0.1)   \\
             &NH$_3$(1,1)     & 1.3(0.3)   & 0.4(0.1)   & 8.1(0.1)   & 1.4(0.3)   \\
             &HCN(1--0)       & 4.6(0.5)   & 1.7(0.2)  & 4.7(0.1) & 1.7(0.2) \\
             &HCO$^+$(1--0)   & 4.3(0.4)   & 2.1(0.2)  & 4.5(0.1) & 2.0(0.2) \\
             &H$^{13}$CN(1--0)& 0.7(0.1)  & 0.27(0.02)  & 7.7(0.1) & 2.0(0.2) \\
             &H$^{13}$CO$^+$(1--0) & 0.26(0.05) & 0.22(0.04)  & 7.4(0.1) & 1.1(0.2) \\
             &CH$_3$OH(2--1)  & 0.27(0.04) & 0.41(0.05)  & 7.98(0.04) & 0.7(0.1) \\
\noalign{\smallskip}\hline
(--100, 20)  &CS(2--1)        &  6.26(0.04) & 4.40(0.04) & 4.97(0.01)  & 1.37(0.01) \\
                &CS(5--4)        &  1.4(0.1) & 1.04(0.06) & 4.95(0.03)  & 1.3(0.1) \\
                &N$_2$H$^+$(1--0)& 12.4(0.1) & 8.5(0.3)$^1$ & 4.663(0.003)& 0.58(0.01) & 4.6(0.6) \\
                &NH$_3$(1,1)     & 10.5(0.2) & 8.7(0.6)$^1$ & 4.625(0.005)& 0.52(0.02) & 3.5(0.6) \\
                &HCN(1--0)       &  8.7(0.5) & 3.4(0.2) & 4.82(0.04) & 1.6(0.1) \\
                &HCO$^+$(1--0)   &  6.3(0.5) & 4.4(0.2) & 4.66(0.04) & 1.4(0.1) \\
                &H$^{13}$CO$^+$(1--0) & 1.0(0.1)& 1.3(0.1)& 4.63(0.02) & 0.67(0.04) \\
\noalign{\smallskip}\hline
(--80, 140)  &CS(2--1)        &  3.0(0.1) & 3.7(0.6) & 5.38(0.01)  & 0.74(0.01) \\
                &CS(5--4)        &           & $<0.5$ \\
                &NH$_3$(1,1)     & 16.0(0.1) & 9.2(0.1)$^1$ & 5.384(0.001)& 0.387(0.004)& 8.5(0.4) \\
\noalign{\smallskip}\hline
(100, --120) &CS(2--1)        &  5.6(0.1) & 3.4(0.1) & 2.04(0.01)  & 1.54(0.03) \\
                &CS(5--4)        &  5.5(0.4) & 3.2(0.2) & 2.04(0.05)  & 1.5(0.1) \\
\noalign{\smallskip}\hline
(0, 0)  &CS(2--1)        &7.64(0.03) & 4.77(0.02)   & 4.91(0.01) & 1.50(0.01) \\
           &CS(5--4)        &  3.2(0.1) & 1.5(0.1)$^2$ & 5.07(0.03) & 1.2(0.1) \\
           &C$^{34}$S(2--1) &  0.7(0.1) & 0.42(0.04)   & 4.92(0.05) & 1.2(0.1) \\
           &N$_2$H$^+$(1--0)&  0.4(0.1) & 0.06(0.01)   & 5.0(0.1)   & 1.7(0.1) \\
           &NH$_3$(1,1)     &  3.4(0.4) & 0.46(0.05)   & 5.4(0.1)   & 2.9(0.4) \\
           &HCN(1--0)       & 12.6(0.3) & 4.4(0.1)     & 4.98(0.02) & 1.7(0.1) \\
           &HCO$^+$(1--0)   &  9.2(0.5) & 4.7(0.2)     & 4.62(0.03) & 1.7(0.1) \\
	   &H$^{13}$CN(1--0)&           & $<0.3$ \\
           &H$^{13}$CO$^+$(1--0) & 0.39(0.04)& 0.27(0.03)& 4.9(0.1)   & 1.4(0.2) \\
\noalign{\smallskip}\hline
(0, 40) &CS(2--1)        &3.52(0.04) & 2.43(0.03) & 5.03(0.01) & 1.39(0.02) \\
           &CS(5--4)        &  0.8(0.1) & 0.45(0.04) & 5.3(0.1)  & 1.8(0.2) \\
           &N$_2$H$^+$(1--0)&           & $<0.1$  \\
           &NH$_3$(1,1)     &  1.6(0.3) & 0.23(0.05)   & 5.0(0.2)   & 2.5(0.6) \\
           &HCN(1--0)       & 12.8(0.4) & 4.5(0.1)     & 5.04(0.02) & 1.6(0.1) \\
           &HCO$^+$(1--0)   &  9.2(0.5) & 4.7(0.2)     & 4.62(0.03) & 1.7(0.1) \\
\noalign{\smallskip}\hline
(0, --40) &CS(2--1)        &6.35(0.04) & 3.25(0.03) & 4.96(0.01)  & 1.89(0.02) \\
             &CS(5--4)        &  2.6(0.1) & 1.41(0.03) & 5.44(0.02)  & 1.7(0.1) \\
             &N$_2$H$^+$(1--0)&           & $<0.1$  \\
             &NH$_3$(1,1)     &  3.1(0.3) & 0.40(0.05)   & 5.3(0.2)   & 2.8(0.4) \\
             &HCN(1--0)       & 10.2(0.4) & 3.1(0.1)     & 5.03(0.03) & 1.8(0.1) \\
             &HCO$^+$(1--0)   &  6.1(0.4) & 3.2(0.2)     & 4.64(0.05) & 1.8(0.1) \\
\noalign{\smallskip}\hline\noalign{\smallskip}
\end{tabular}

\scriptsize

\flushleft{
$^1$ -- the $T_{\rm EX}$ value of the main group of hyperfine components \\
$^2$ -- the spectrum possesses two overlapping components, the second one of lower intensity at $\sim 6.7$~km~s$^{-1}$.
}

\label{mol_peaks}
\end{table*}

\smallskipamount 3pt

\begin{figure*}

\begin{minipage}[b]{0.49\textwidth}
  \includegraphics[width=\textwidth]{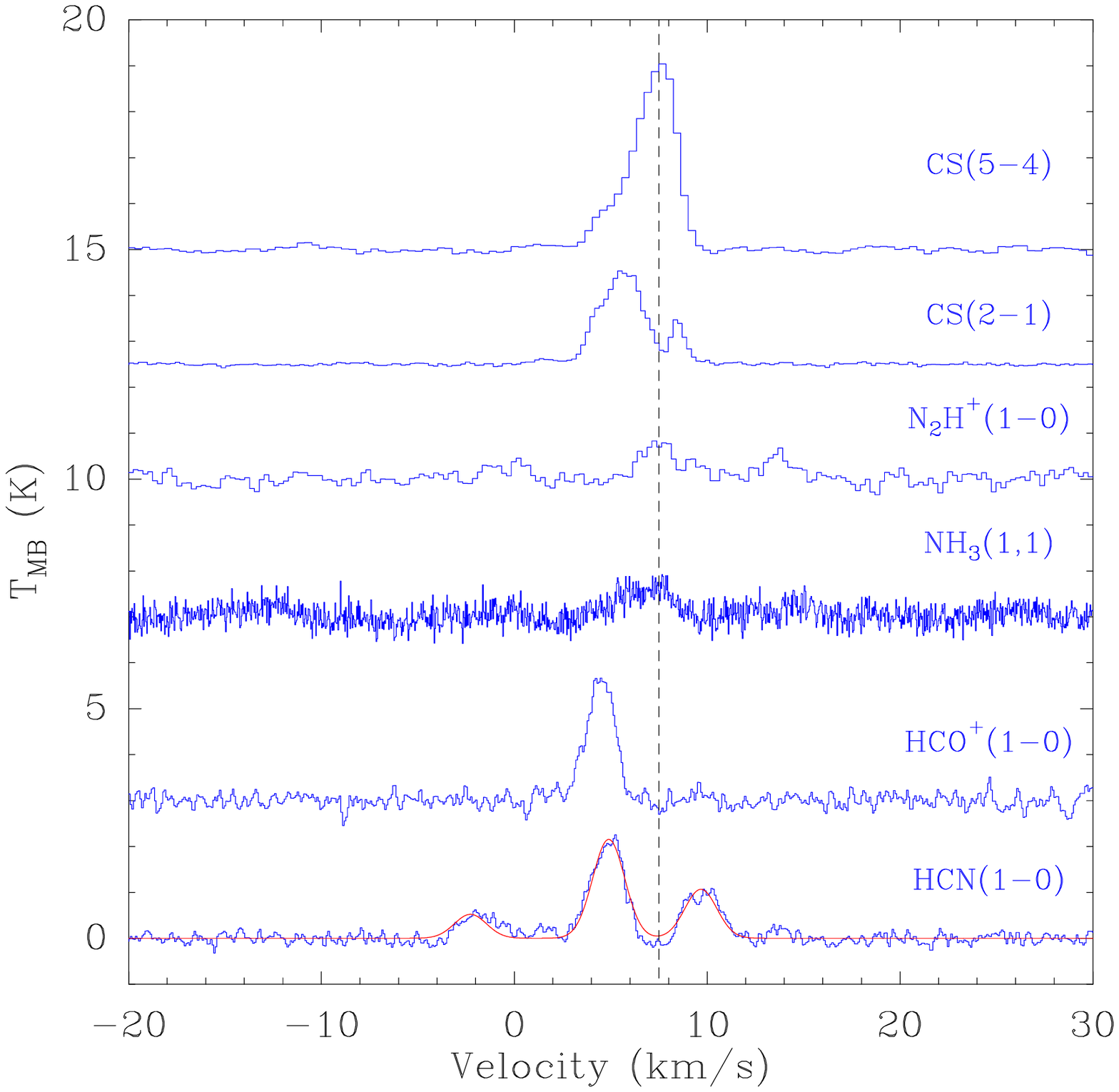}
\end{minipage}
\hspace{0.5mm}
\begin{minipage}[b]{0.46\textwidth}
  \includegraphics[width=\textwidth]{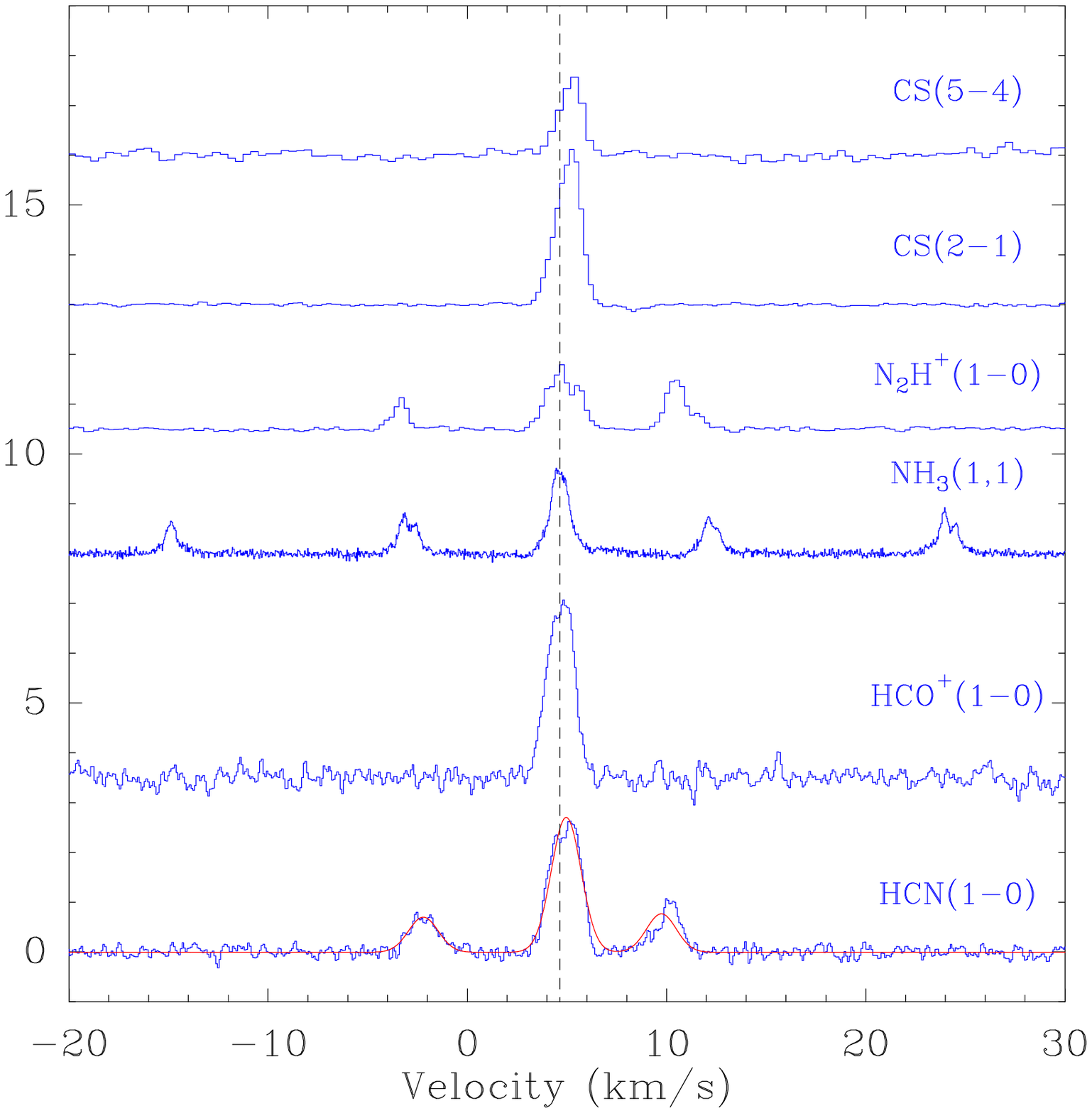}
\end{minipage}

\caption{\small
Molecular line spectra averaged over the eastern (left panel)
($\Delta\alpha=100''...40''$, $\Delta\delta=-50''...60''$)
and the western (right panel) parts
($\Delta\alpha=-50''...-140''$, $\Delta\delta=-50''...60''$)
of the region observed.
Dashed vertical lines correspond to 7.5~km~s$^{-1}$ (left panel)
and to 4.65~km~s$^{-1}$ (right panel).
The results of the HCN(1--0) fitting by Gaussian triplet
of the same width are also shown by smooth curve.
}
\label{spectra}
\end{figure*}

Towards the eastern part of the region the observed lines
can be roughly divided into two velocity ranges.
The HCN(1--0) and HCO$^+$(1--0) lines are observed
at $\sim 4.5-5$~km~s$^{-1}$.
Other lines are observed at $\sim 7-8$~~km~s$^{-1}$.
The averaged CS(5--4), CS(2--1), N$_2$H$^+$(1--0), NH$_3$(1,1), HCO$^+$(1--0)
and HCN(1--0) spectra for the eastern and western areas
are shown on left and right panels of Fig.~\ref{spectra}, respectively.
The HCN(1--0) fitting results are also shown on both panels.
The CS(2--1) profiles in the eastern area are self-reversed.
Both CS lines in the eastern area have extended blue wings
probably due to emission at $\sim 4.5-5$~km~s$^{-1}$.
Weak H$^{13}$CN(1--0) and H$^{13}$CO$^+$(1--0) emission
is detected towards distinct eastern positions
at $\sim 7.3-7.5$~km~s$^{-1}$ (Table~\ref{mol_peaks}).
No HCN(1--0) or HCO$^+$(1--0) emission has been detected at these
velocities.
There is a hint of weak absorption
on the HCN(1--0) spectra at $\sim 7.5$~km~s$^{-1}$ in the eastern area
(Fig.~\ref{spectra}, left panel).
We discuss briefly this phenomenon
in Section~\ref{chemistry}.

The line widths in different parts of the cloud
lie in the range: $\sim 1-2$~km~s$^{-1}$ (Table~\ref{mol_peaks})
being much higher than thermal widths
(ammonia kinetic temperatures and dust temperatures are given
in Tables~\ref{masses} and \ref{LTE}, respectively).
Enhanced line widths of weak ammonia lines towards
the central and eastern positions
are probably due to systematic uncertainties
of the fits for the spectra with low signal-to-noise ratios.
The broadening of the CS(5--4) and C$^{34}$S(2--1) lines
towards $(80'',40'')$ can be connected with enhanced
dynamical activity and the presence of systematic motions
(Section~\ref{modeling}).
Towards the $(-100'',20'')$ position close to the center
of the dust clump~6 the N$_2$H$^+$, NH$_3$ and H$^{13}$CO$^+$
lines are more narrow ($0.5-0.7$~km~s$^{-1}$) than
the CS, HCN and HCO$^+$ lines ($1.3-1.6$~km~s$^{-1}$).
This can be due both to optical depth effects
and to the fact that the lines trace different gas components
on the line of sight with different extent of turbulence,
e.g. the quiescent region within the clump
and the surrounding gas with higher extent of turbulence, respectively.
Most narrow lines are detected towards the $(-80'',140'')$ position
(ammonia clump~2) implying the lowest extent of turbulence.

\subsection{Comparison with CO data}
\label{CO}

\begin{figure}
    \includegraphics[width=0.45\textwidth,angle=-90]{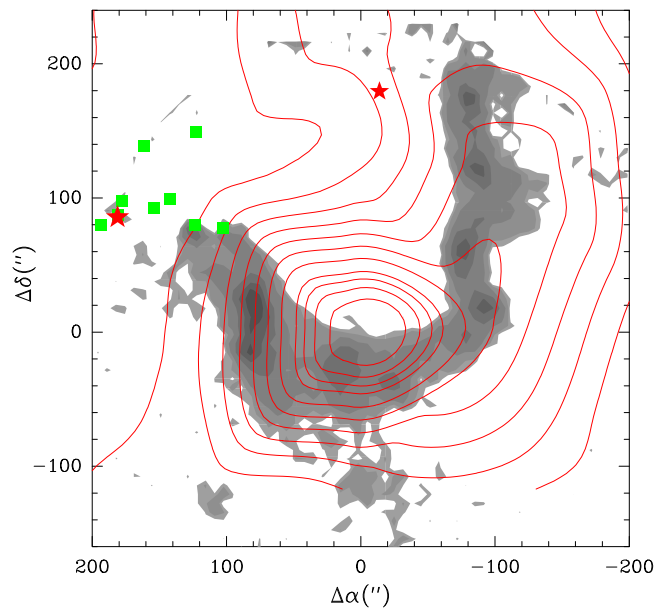}

\caption{\small
The $^{13}$CO(3--2) integrated intensity map observed by Zhu et al. (2006)
(red contours) is overlayed over 1.2~mm dust continuum emission (greyscale).
Intensity contours range from 10\% to 90\% with the step of 10\% of the peak
value (39.5~K~km~s$^{-1}$) plus the contour of 95\%.
The rest of symbols are the same as in Fig.~\ref{line_maps}.
}
\label{13CO_map}
\end{figure}

The W40 molecular cloud has been mapped previously in the CO isotopic lines
by different authors (Zeilik \& Lada 1978, Crutcher \& Chu 1982,
Vall\'ee et al. 1992, Zhu et al. 2006).
The $^{13}$CO(3--2) map from Zhu et al. (2006) overlayed over
our dust continuum map is shown in Fig.~\ref{13CO_map}.
They also found implication for the weak CO outflow in the center of the map.
No corresponding dust or molecular emission peaks are found
near this position in our observations.
The shape of the $^{13}$CO(3--2) map in the western part
is curved to the north correlating in general with
the shape of the dust ring, yet, the spatial resolution ($\sim 80''$)
is low to resolve the structure of emission region in details.
The CO emission is also detected towards the center of the H~II region
and to the east from the center.
The CO and $^{13}$CO spectra consist of the main component
at $\sim 4.5-5$~km~s$^{-1}$ and weak satellites
in emission and in absorption at higher velocities.
Near the center of the H~II region an additional component
at $\sim 9-10$~km~s$^{-1}$ with intensity of the order
of the main component appears on the spectra.
For three central positions of our region with no
kinetic or dust temperature estimates
we took the CO(1--0) peak main beam temperatures as crude estimates
of kinetic temperature in model calculations (Section~\ref{Non-LTE}).

\section{Physical parameters derived from mm-continuum data}

In Table~\ref{masses} physical parameters of the dust clumps are given.
Dust temperatures ($T_{\rm d}$) are taken from Maury et al. (2011)
for the clumps associated with their sources.
The eastern clumps which are closer to the cluster sources
have higher dust temperatures compared to the southern and western clumps.
For the remaining clumps dust temperatures are set to 20~K.
Dust masses are calculated as (e.g. Doty \& Leung 1994):

\begin{equation}
M_{\rm d} =\frac{F_{\rm total}\,D^2}{k_{1.2}\,B_{1.2}(T_{\rm d})}~,
\end{equation}

{\noindent where $D$, $k_{1.2}$ and $B_{1.2}(T_{\rm d})$ are the
source distance, the dust mass absorption coefficient and Planck function
at 1.2~mm, respectively.}
For circumstellar envelopes around Class~I and Class~0 protostars
the mass absorption coefficient can be taken equal to 1~cm$^2$ g$^{-1}$
(Motte et al. 1998), the value at 1.3~mm wavelength appropriate
for cold dust grains covered with thick icy mantles (Ossenkopf \& Henning 1994).
We used this value in calculations.
For prestellar dense clumps Motte et al. (1998) adopted
$k_{1.2}=0.5$~cm$^2$ g$^{-1}$.
If we take this value for the sourceless clumps 5 and 8
their masses, column densities and number densities will be two times higher.
Gas masses are calculated as $M=R_m\,M_{\rm d}$.
A gas-to-dust mass ratio, $R_m$, is taken equal to 100.
The uncertainties of masses ($\sim 30-35$\%) are calculated using propagation
of errors method using the uncertainties of fluxes,
dust temperatures uncertainties ($\sim 20$\% according to Maury et al. 2011)
and distance ($\sim 10$\% according to Shuping et al. 2012).

Peak hydrogen column densities ($N_{\rm H_2}$) have been calculated from
peak flux-per-beam and dust temperature values (e.g. Motte et al. 1998):

\begin{equation}
N_{\rm H_2}=\frac{F_{\rm peak}}{\Omega\,m\,R_m^{-1}\,k_{1.2}\,B_{1.2}(T_{\rm d})}~,
\label{nh2}
\end{equation}

{\noindent where $\Omega$ is the beam solid angle, $m=2.33~amu$
is mean molecular mass.}
The $N_{\rm H_2}$ positions are located close
to the positions of clump centers (Table~\ref{dust_clumps}).
The $N_{\rm H_2}$ values are given in column 4 of Table~\ref{masses}.
The calculated uncertainties of the $N_{\rm H_2}$ values are $\sim 25-35$\%.
Besides, we have calculated mean volume densities
for the individual clumps areas ($\bar n$) as mean column density
divided by the clump size (Table~\ref{dust_clumps}).
They are given in the last column of Table~\ref{masses}.

\begin{table}
\centering
\caption[]{Physical parameters derived from 1.2 mm continuum data}
\small
\begin{tabular}{lrrrr}
\noalign{\hrule}\noalign{\smallskip}
Object   & $T_{\rm d}$ (K) & $M (M_{\odot}$)
         & $N_{\rm H_2}$ (cm$^{-2}$) & $\bar n$ (cm$^{-3}$) \\
\noalign{\smallskip}\hline\noalign{\smallskip}

Clump 1     & 36$^a$ &   0.4       & 2.5\,10$^{22}$  & 3.1\,10$^5$ \\
Clump 2     & 27$^a$ &   8.1$^c$   & 6.2\,10$^{22}$  & 1.8\,10$^5$ \\
Clump 3     & 28$^a$ &             & 5.2\,10$^{22}$  & 6.6\,10$^5$ \\
Clump 4     & 20$^b$ &   5.6       & 5.2\,10$^{22}$  & 1.0\,10$^5$ \\
Clump 5     & 20$^b$ &   2.2       & 4.3\,10$^{22}$  & 1.4\,10$^5$ \\
Clump 6     & 18$^a$ &   3.6       & 7.0\,10$^{22}$  & 2.9\,10$^5$ \\
Clump 7     & 20$^b$ &   2.2       & 7.3\,10$^{22}$  & 2.8\,10$^5$ \\
Clump 8     & 13$^a$ &   6.2       & 1.1\,10$^{23}$  & 3.6\,10$^5$ \\
Clump 9     & 23$^a$ &   2.0       & 6.2\,10$^{22}$  & 1.4\,10$^6$ \\
\noalign{\smallskip}\hline\noalign{\smallskip}
\end{tabular}

\small
\flushleft{
$^a$ -- dust temperatures taken from Maury et al. (2011) \\
$^b$ -- assumed dust temperature  \\
$^c$ -- total mass for clumps 2 and 3
}
\label{masses}
\end{table}

\section{Physical parameters derived from molecular line data}

Molecular line data are used to derive physical parameters of the gas
(kinetic temperatures, densities, column densities, abundances
and virial masses) using either the LTE approximation or non-LTE modeling.
For most positions we assume that the dust coexists with the gas traced
by molecular lines.
For the positions with no kinetic or dust temperature estimates
we used the CO(1--0) line peak temperatures as a measure of kinetic temperature.
Yet, the CO line velocities usually slightly differ from velocities
of the other lines.

\subsection{Kinetic temperatures and column densities derived from LTE analysis}


We derived NH$_3$ column densities in the LTE approximation
using the procedures from Harju et al. (1993).
For the western positions where ammonia lines are strong
we used the (1,1) line width, total optical depth
and excitation temperature as well as
integrated intensity of the optically thin (2,2) line in the calculations.
Kinetic temperatures are calculated from rotational temperatures
using an empirical relation from Tafalla et al. (2004).
The $N$(NH$_3$) and $T_{\rm KIN}$ values for the positions close
to the centers of ammonia clumps are given in Table~\ref{LTE}.
The uncertainties of kinetic temperatures are estimated
by the propagation of errors method using uncertainties
of integrated intensities and uncertainties of fits.
The uncertainties of column densities are $\sim 10-30$\%.
The kinetic temperature of the ammonia clump~1 is
close to the dust temperature of the associated dust clump~6
(Table~\ref{masses}).
The kinetic temperature of the ammonia clump~2 is higher than
the dust temperature of the dust clump~8
and is lower than of the dust clump~9.
Kinetic temperature map has been obtained for the ammonia clump~2.
It shows no apparent spatial variations.
The NH$_3$ column densities for the
eastern and central positions
are calculated from the (1,1) integrated intensities
in the LTE and optically thin approximation
taking excitation temperature equal to 10~K.
The $N$(NH$_3$) uncertainties for these positions
are connected mainly with unknown excitation temperature
and can reach $\sim 60$\%.

\begin{table*}
\centering
\caption[]{Ammonia kinetic temperatures and LTE column densities}
\small
\begin{tabular}{rrrrrrrr}
\noalign{\hrule}\noalign{\smallskip}
($\Delta\alpha$, $\Delta\delta$) & $T_{\rm KIN}$ & $N$(NH$_3$)  & $N$(N$_2$H$^+$) & $N$(H$^{13}$CN) & $N$(H$^{13}$CO$^+$) &  $N$(C$^{34}$S) & Associated \\
  $('', '')$                       &    (K)        &  (cm$^{-2}$) &  (cm$^{-2}$)    &  (cm$^{-2}$)    &  (cm$^{-2}$)        &   (cm$^{-2}$) & objects  \\
\noalign{\smallskip}\hline\noalign{\smallskip}

(80, 40)    &           & $6.0\,10^{13}$ & 7.3\,10$^{11}$ & 1.4\,10$^{12}$ &                & $2.1\,10^{13}$ & Dust clump 2    \\
(80, 0)     &           & $5.2\,10^{13}$ & 1.3\,10$^{12}$ & 1.1\,10$^{12}$ & 2.4\,10$^{11}$ & $1.1\,10^{13}$ & Dust clump 3    \\
(0, 0)      &           & $8.4\,10^{13}$ & 4.2\,10$^{11}$ &                & 3.6\,10$^{11}$ & $2.9\,10^{12}$ & $^{13}$CO peak  \\
(0, --40)   &           & $8.1\,10^{13}$ \\
(0, 40)     &           & $5.0\,10^{13}$ \\
(--100, 20) & 20.7(2.5) & $2.5\,10^{14}$ & 1.5\,10$^{13}$ &                & 8.3\,10$^{11}$ &                & Dust clump 6   \\
(--80, 140) & 15.9(0.4) & $5.7\,10^{14}$ &                &                &                &                & Ammonia clump 2 \\

\noalign{\smallskip}\hline\noalign{\smallskip}
\end{tabular}

\label{LTE}
\end{table*}

The N$_2$H$^+$(1--0) line towards the $(-100'',20'')$ position
is optically thick and has an excitation temperature close
to the NH$_3$(1,1) one (Table~\ref{mol_peaks}).
The H$^{13}$CO$^+$ column density for this position
is calculated with the same value of excitation temperature.
The H$^{13}$CN, H$^{13}$CO$^+$ and N$_2$H$^+$ column densities
towards the eastern and central positions
are calculated in the LTE and optically thin approximation
taking $T_{\rm EX}=10$~K.
The $N$(C$^{34}$S) values are calculated
for three selected positions under the same assumptions.
The C$^{34}$S excitation temperature for the $(80'',40'')$ position
is taken equal to 50~K according to the results of non-LTE model calculations
(Section~\ref{modeling}).
For the $(80'',0'')$ and $(0'',0'')$ positions the C$^{34}$S excitation temperatures
are taken equal to 30~K and 10~K, respectively.
The calculated values are given in Table~\ref{LTE}.

\subsection{Physical parameters derived from non-LTE analysis}
\label{Non-LTE}

Knowing kinetic temperature, molecular column density and
peak molecular line intensity and width it is possible
to derive number density using non-LTE (LVG or microturbulent)
approach for homogeneous isothermal model.
If two different transitions of the same molecule are observed,
as in the case of CS, one can get molecular column density and number density
from the observed peak line intensities and widths.

We used RADEX non-LTE radiative transfer online code (Van der Tak et al. 2007)
and derived number densities and CS column densities towards selected positions
where the CS line profiles are close to Gaussian.
Kinetic temperatures are taken close to the
ammonia (Table~\ref{LTE}) or dust temperatures (Table~\ref{masses}).
For the $(0'',0'')$ and $(0'',\pm 40'')$ positions
kinetic temperatures estimates are taken
close the CO(1--0) main beam temperatures.
In order to derive parameters towards one
selected eastern position where the CS(2--1) line profile is non-Gaussian
and self-absorbed we performed model calculations in the framework
of inhomogeneous non-LTE model (Section~\ref{modeling}).

Having calculated number densities we used RADEX
to get column densities of other species assuming they coexist with CS.
The results are given in Table~\ref{densities}.
The derived column densities of the optically thin H$^{13}$CO$^+$(1--0)
and N$_2$H$^+$(1--0) lines are practically
the same as those derived from the LTE analysis (Table~\ref{LTE})
confirming the validity of the LTE approximation.
The upper limit on number density and the lower limit on the CS column density
are derived for the $(-80'',140'')$ position where
only upper limit on the CS(5--4) line temperature is measured.
Towards the $(120'',60'')$ position (dust clump~1)
the CS(2--1) spectrum consists of two peaks (Table~\ref{mol_peaks})
while the CS(5--4) profile is single-peaked.
The CS(5--4) and CS(2--1) (red peak) velocities
are 8.3 and 8.9~km~s$^{-1}$, respectively.
The physical parameters derived on a base of these lines parameters
should be treated with caution.
The derived number densities
are higher than mean densities of the associated dust clumps
(Table~\ref{masses}).

Taking the N$_2$H$^+$ emission region size and the N$_2$H$^+$ line width
we have calculated virial mass for the N$_2$H$^+$ clump
(associated with the ammonia clump~1 and the dust clump~6)
assuming it to be spherically-symmetric
with no external pressure and no magnetic field:
$M_{\rm VIR}=105\,\Delta V^2\,(1.2 d)$.
The coefficient 1.2 is used to convert the FWHM size
to the size at the e$^{-1}$ level to be compared with mass estimated
from the 1.2~mm continuum data.
The derived virial mass, $\sim 3$~$M_{\odot}$, is in general agreement with
the mass derived from the continuum data (Table~\ref{masses})
taking into account possible uncertainties.
Virial mass for the ammonia clump~2 $(\sim 1.9$~$M_{\odot}$) derived from
the ammonia data is close to the mass of the dust clump 9.

\begin{table*}
\centering
\caption[]{The parameters derived from the non-LTE modeling}
\small
\begin{tabular}{rrrrrrr}
\noalign{\hrule}\noalign{\smallskip}
($\Delta\alpha$, $\Delta\delta$)
                         & $T_{\rm KIN}$  & $n$          & $N$(CS)        & $N$(HCO$^+$) & Associated \\
 $('', '')$              &    (K)         & (cm$^{-3}$)  & (cm$^{-2}$)    & (cm$^{-2}$)  & object \\
\noalign{\smallskip}\hline\noalign{\smallskip}

 (120, 60)   & 36             & $3.2\,10^6$\,$^a$  & $6.3\,10^{12}$\,$^a$ &     & Dust clump 1        \\
 (80, 40)    & 40             & $1.5\,10^6$\,$^b$  & $1.7\,10^{14}$       &     & Dust clump 2        \\
 (0, 0)      & 25             & $5.0\,10^5$  & $2.9\,10^{13}$ & $9.0\,10^{12}$ & $^{13}$CO peak      \\
 (0, --40)   & 20             & $1.0\,10^6$  & $3.0\,10^{13}$ & $7.0\,10^{12}$ &                     \\
 (0, 40)     & 25             & $3.3\,10^5$  & $1.5\,10^{13}$ & $8.5\,10^{12}$ &                     \\
 (--100, 20) & 20             & $5.1\,10^5$  & $3.0\,10^{13}$ & $6.8\,10^{12}$ & Dust clump 6        \\
 (--80, 140) & 16             & $\la\,10^5$  & $\ga 1.5\,10^{13}$         &     & Ammonia clump 2     \\
 (100, --120)& 26             & $1.8\,10^6$  & $3.0\,10^{13}$             &     & Southern CS clump     \\

\noalign{\smallskip}\hline\noalign{\smallskip}
\end{tabular}

\small
\flushleft{
$^a$ -- due to the difference between the CS(2--1) and (5--4) velocities
these estimates are questionable. \\
$^b$ -- the density for the core center,
the density in the envelope is $1.5\times 10^3$~cm$^{-3}$
(Section~\ref{modeling}).
}

\label{densities}
\end{table*}

\subsection{Non-LTE modeling of CS spectra with inhomogeneous model}
\label{modeling}

\begin{figure}
    \includegraphics[width=0.45\textwidth]{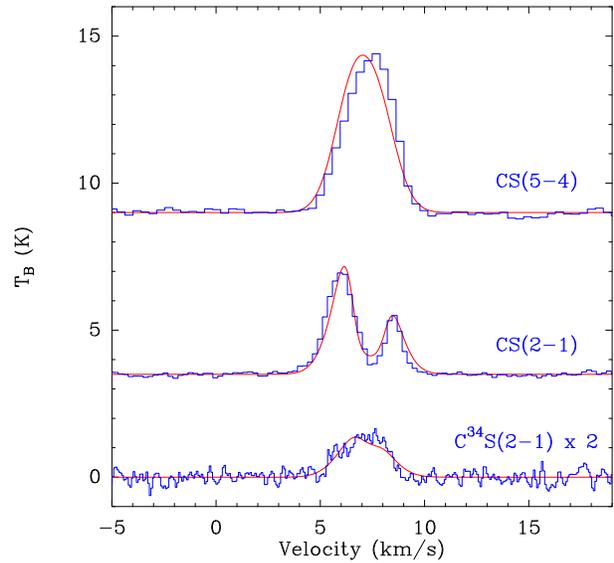}
\caption{
The observed CS(5--4), CS(2--1) and C$^{34}$S(2--1) line profiles
towards the $(80'',40'')$ position.
The results of model calculations in the framework
of inhomogeneous microturbulent model with systematic motions
are shown as red smooth curves (see text for details).
}
\label{80_40_cs}
\end{figure}

Towards the eastern branch of the dust ring
(the area that includes the dust clumps~2 and 3)
the CS(5--4) lines are the most intensive and single-peaked while
the CS(2--1) profiles are asymmetric and self-absorbed.
The CS line profiles towards $(80'',40'')$ (Fig.~\ref{80_40_cs})
are in less extent affected by the emission at lower velocities
compared with other positions in this area.
We used them for modeling.

The velocity of the absorption dip on the CS(2--1) profile is close
to the peak of the C$^{34}$(2--1) profile which is probably optically thin
(Fig.~\ref{80_40_cs}).
This implies that the observed CS(2--1) profile asymmetry
is connected with systematic motions on the line of sight.
The type of asymmetry with blue peak stronger than the red one
can be a result of the infall motions on the line of sight
as predicted by the ``inside-out'' model of low-mass star formation
(Shu 1977, Evans 1999).
The signatures of infall on line profiles are widely observed
both in low-mass and high-mass star forming regions
(e.g. Lee et al. 2001, Sohn et al. 2007, Klaassen \& Wilson 2007, Wu et al. 2007).
If the gas on the line of sight is contracting
a crude estimate of the infall velocity, $V_{\rm IN}$,
can be made with the 2-layer model (Myers et al. 1996).
We calculated the $V_{\rm IN}$ value for the CS(2--1) profile
of $\sim 0.24$~km~s$^{-1}$.
The kinematic mass infall rate can be estimated as
$dM/dt=4\pi\,r_{\rm in}^2\,m\,n_0\,V_{\rm IN}$
(Myers et al. 1996, Klaassen \& Wilson 2007),
where $r_{\rm in}$ is the radius of the infall region,
$m$ is mean molecular mass, $n_0$ is an ambient density.
Taking $r_{\rm in}\sim 0.05$~pc (20$''$ at distance of 500~pc)
and $n_0\sim 1.5\times 10^3$~cm$^{-3}$ (the density of envelope, see below)
the crude estimate for the mass infall rate
is $\sim 2.7\times 10^{-6}$~$M_{\odot}$~yr$^{-1}$.
This is much lower than the mass infall rates calculated
for HMSF regions (e.g. Klaassen \& Wilson 2007, Fuller et al. 2005)
or massive YSOs (Chen et al. 2010)
being more in accordance with the values for low-mass starless cores
(Lee et al. 2001).
Note, that we used an arbitrary value of $r_{\rm in}$ in calculations.

In order to get other physical parameters
of molecular gas towards the $(80'',40'')$ position
we performed model calculations of CS line excitation
and fit the model profiles to the observed ones.
The microturbulent multi-layer spherically-symmetric model
has been used.
Layer widths are scaled logarithmically and physical
parameters are the functions of layer radius.
In addition to the number of logarithmically scaled layers (core)
an outer layer (envelope) with constant density is added
in order to allow absorption of the CS(2--1) emission in low-density gas.
The method of calculations is similar to the one described
by Turner et al. (1997) (see their Appendix).
In order to take into account systematic velocity field
each layer is divided into the number of steps at which
the projection of systematic velocity is calculated.
Systematic velocity is set to zero in the central layer.
Number density and systematic velocity fall down outwards according
to the power-law radial profiles.
Kinetic temperature, turbulent velocity dispersion
and CS abundance are set constant throughout the cloud.
We used the CS--H$_2$ collisional rates tabulated
for given kinetic temperatures by Turner et al. (1992).

The results of model calculations are shown in Fig.~\ref{80_40_cs}
as smooth curves overlayed over the observed line profiles.
The model parameters are the following:
$T_{\rm KIN}=40$~K, $V_{\rm IN}=0.5~(r/R_0)^{-0.1}$~km~s$^{-1}$,
$R_{\rm core}/R_0=33$, $R_{\rm env}/R_{\rm core}=40$,
where $R_0$, $R_{\rm core}$, and $R_{\rm env}$
are radii of the central layer, the core and the envelope, respectively.
Total column densities are $N$(CS)=$1.7\times 10^{14}$~cm$^{-2}$
and $N$(C$^{34}$S)=$2\times 10^{13}$~cm$^{-2}$.
Densities and turbulent velocity dispersions in the core
and in the envelope are:
$n_{\rm core}(r)=1.5\times 10^6 (r/R_0)^{-2}$~cm$^{-3}$,
$n_{\rm env}=1.5\times 10^3$~cm$^{-3}$,
$V_{\rm core}=1.4$~km~s$^{-1}$ and $V_{\rm env}=0.7$~km~s$^{-1}$.
The envelope effectively absorbs the CS(2--1) emission producing
self-adsorption dip on the line profile while the presence
of infall motions makes the CS(2--1) profile asymmetric.
The $V_{\rm IN}$ value at the boundary of the cloud
is $\sim 0.25$~km~s$^{-1}$ in accordance
with the crude estimate of the 2-layer model (Myers et al. 1996).
Total CS column density and number density in the center
of the core derived from the model are given in Table~\ref{densities}.

The peak on the observed CS(5--4) profile is shifted
by $\sim 0.2$~km~s$^{-1}$ to the higher velocities with respect
to the peak of the C$^{34}$S(2--1) profile
and the self-absorption dip on the CS(2--1) profile.
If this is connected with an absorption of the CS(5--4) blue wing
in the foreground gas
(which is probably seen on the C$^{34}$S(2--1) profile)
the observed CS(5--4) intensities could be underestimated and
the central density in the core could be higher.

\subsection{Molecular and electron abundances}

Molecular abundances are given in Table~\ref{abundances}.
They are calculated as the ratios of molecular column densities
(Tables~\ref{LTE} \& \ref{densities})
to the H$_2$ column densities ($N_{\rm MOL}/N$(\rm H$_2$)).
The latter have been calculated using the equation~\ref{nh2}
for the areas with a 20$''$ radius around selected positions
using dust temperature or kinetic temperature estimates
(Table~\ref{densities}).
No abundance estimates are done towards the $(0'',40'')$ position where
no noticeable dust continuum emission is detected
and dense gas and dust apparently do not coexist there.
We have not estimated abundances for the $(0'',0'')$ position
where coexistence of gas and dust towards is also questionable
(Fig.~\ref{line_maps}).

Molecular abundances strongly vary throughout the cloud.
The abundances of nitrogen-bearing molecules
(NH$_3$ and N$_2$H$^+$) and of H$^{13}$CO$^+$ are enhanced
towards western positions
while the CS abundances are enhanced towards the eastern position
$(80'',40'')$ (dust clump 2) and in the southern CS clump.

Taking HCO$^+$ abundance and number density it is possible to estimate
electron abundance, $X$(e), using relation that follows from
equations of formation and destruction of the HCO$^+$ and H$_3^+$ ions
in dense interstellar clouds (e.g. Zinchenko et al. 2009):

\begin{equation}
X({\rm HCO^+})=\frac{\zeta/n}{\alpha({\rm HCO^+})\,X(e) + k_g\,X_g}~.
\end{equation}

{\noindent We assumed the cosmic-ray ionization rate to be equal
to $\zeta=3\times 10^{-17}$~s$^{-1}$ while
the HCO$^+$ dissociative recombination rate, $\alpha({\rm HCO^+})$,
is assumed to be equal to $7.5\times 10^{-7}$~s$^{-1}$~cm$^3$ (Turner 1995).}
The HCO$^+$ recombination onto negatively charged dust grains
(the term $k_g\,X_g$, where $k_g$ and $X_g$ are
the rate coefficient and the abundance of dust grains, respectively)
is much lower compared
with cosmic-ray ionization rate divided by the product
of number density and HCO$^+$ abundance and can be neglected
(Zinchenko et al. 2009).
The calculated electron abundances are given in Table~\ref{abundances}.
They are among the highest values of $X$(e)
derived both for low-mass (Caselli et al. 1998)
and HMSF (Zinchenko et al. 2009) regions.
The short discussion on the derived abundances is given
in Section~\ref{chemistry}.

\begin{table*}
\centering
\caption[]{Molecular and electron abundances}
\small
\begin{tabular}{rllllllll}
\noalign{\hrule}\noalign{\smallskip}
($\Delta\alpha$, $\Delta\delta$)  & $X$(NH$_3$)   & $X$(N$_2$H$^+$)  & $X$(CS)          & $X$(C$^{34}$S)  & $X$(HCO$^+$)    & $X$(H$^{13}$CO$^+$)& $X$(H$^{13}$CN)& $X$(e) \\
 $('', '')$  \\
\noalign{\smallskip}\hline\noalign{\smallskip}

 (120, 60)          &                &                 &  $8.3\,10^{-10}$                   \\
 (80, 40)           & $2.5\,10^{-9}$ & $3.0\,10^{-11}$ &  $7.1\,10^{-9} $ & $8.3\,10^{-10}$ &                 &                    & $5.8\,10^{-11}$ \\
 (80, 0)            & $1.5\,10^{-9}$ & $3.8\,10^{-11}$ &                  & $3.2\,10^{-10}$ &                 &$7.1\,10^{-12}$     & $3.2\,10^{-11}$ \\
 (0, --40)          & $2.6\,10^{-9}$ &                 &  $9.7\,10^{-10}$ &                 & $2.3\,10^{-10}$ &                    &                &$1.7\,10^{-7}$ \\
 (--100, 20)        & $9.3\,10^{-9}$ & $5.6\,10^{-10}$ &  $1.1\,10^{-9}$  &                 & $2.5\,10^{-10}$ &$3.1\,10^{-11}$     &                &$3.1\,10^{-7}$ \\
 (--80, 140)        & $1.5\,10^{-8}$ &                 &  $>4\,10^{-10}$                  \\
 (100, --120)       &                &                 &  $8.3\,10^{-9}$                    \\

\noalign{\smallskip}\hline\noalign{\smallskip}
\end{tabular}


\label{abundances}
\end{table*}

\section{Morphology of the ionized gas}

\begin{figure*}

\begin{minipage}[b]{0.485\textwidth}
  \includegraphics[width=\textwidth,angle=-90]{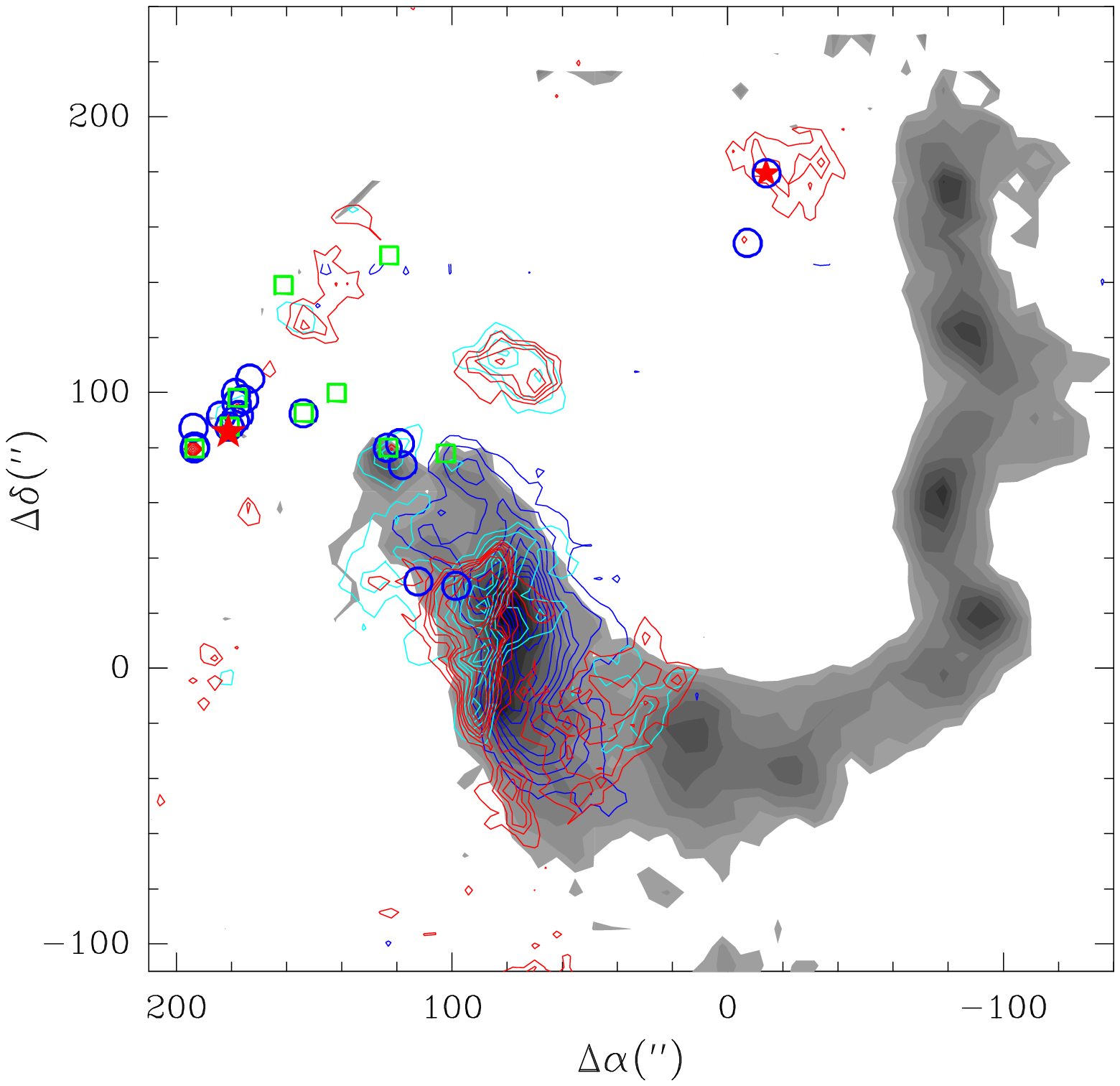}
\end{minipage}
\hspace{2mm}
\begin{minipage}[b]{0.485\textwidth}
  \includegraphics[width=\textwidth,angle=-90]{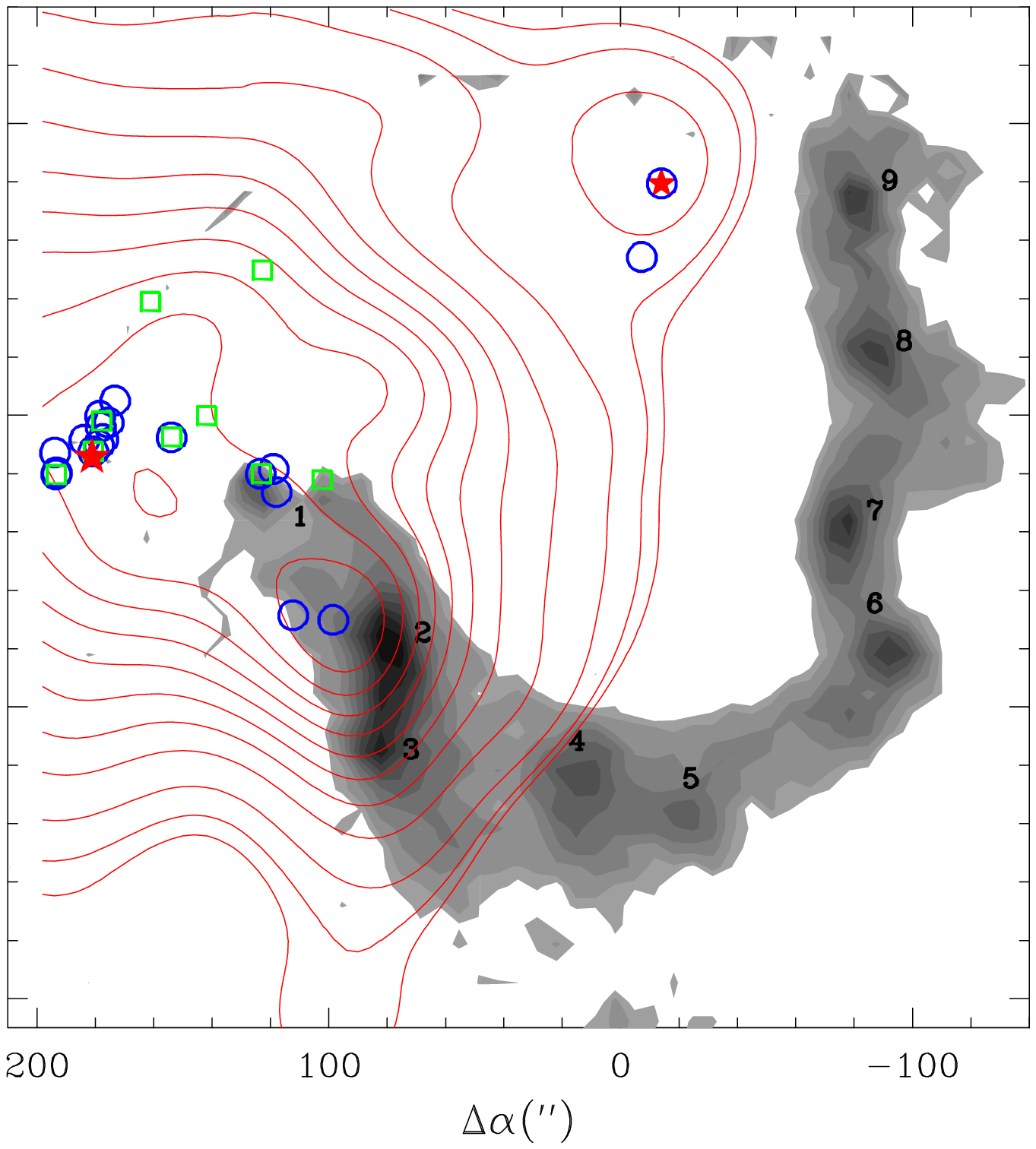}
\end{minipage}

\caption{\small
Left panel: the high-resolution 1280~MHz and 610~MHz maps
(spatial resolutions $\sim 2.4''$ and $\sim 5''$, respectively)
and the CS(5--4) integrated intensity map
(red, cyan and blue contours, respectively).
The 1280~MHz contours are 0.5, 0.7, 0.9, 1.1, 1.5, 2, 3, 4 and 5~mJy~beam$^{-1}$.
The 610~MHz contours range from 3 to 10~mJy~beam$^{-1}$
with 1~mJy~beam$^{-1}$ step.
The CS(5--4) contours range from 10\% to 90\% of the peak value
(19.9~K~km~s$^{-1}$) with 10\%~step.
Right panel: the low-resolution 1280~MHz map ($\sim 45''$)
at 3\%, 5\% and from 10\% to 90\% of peak flux (449~mJy~beam$^{-1}$)
with the 10\% step (Mallick et al. 2013).
The dust clumps are marked by numbers on the right panel.
The contour maps on both panels are
overlayed over the dust continuum image (greyscale).
The rest of symbols are the same as in Fig.~\ref{line_maps}.
}
\label{ionized_gas}
\end{figure*}

Low-resolution 1280~MHz and 610~MHz images of W40 H~II region
derived from our GMRT data are given in Mallick et al. (2013).
The H~II region appears to be bounded to the west
of the main ionizing source (IRS~1A South) by dense gas and dust.
We used high resolution GMRT maps to study an interaction region
between ionized and dense neutral gas and dust in detail.
The 1280 MHz and 610 MHz maps
and the CS(5--4) intensity map integrated in the (6...9)~km~s$^{-1}$
velocity range are shown in Fig.~\ref{ionized_gas} (left panel).
The maps are overlayed over dust continuum image.

We detected an area of enhanced intensity
close to the eastern branch of the dust ring.
This area was observed before at 1465~MHz
by Vall\'ee \& MacLeod (1991).
Our interferometric maps reveal complex clumpy structure of the ionized gas
in this area.
The shapes of ionized, molecular and dust maps are similar in general.
Yet, the ionized gas region is slightly shifted (by several arcseconds)
to the east with respect to the dust and the CS(5--4) emission regions.
Low resolution image shows that ionized gas emission
extends to the west of the cluster, up to the central position
on our maps ($^{13}$CO peak) and includes compact area around IRS~5
at the north-west (Fig.~\ref{ionized_gas}, right panel).
There is another local emission subregion on the 1280~MHz low resolution map
close to the dust clump~2 (Fig.~\ref{ionized_gas}, right panel)
which could also be a Str\"omgren sphere around young massive star.
The VLA~3 source with the 2MASS, X-ray counterparts is located
in the center of this subregion.
The Class~I source is closely located (Fig.~\ref{dust_sources}).
We calculated fluxes of the enhanced intensity area
($\Delta\alpha=120''...70''$, $\Delta\delta=-40''...50''$)
which are: $F(1280)=502(1)$~mJy and $F(610)=555(2)$~mJy.
The spectral index ($\sim -0.14$)
is close to the value expected for optically thin free-free emission.


\section{Discussion}
\label{discussion}


\subsection{The clumpy ring}

The observed in the 1.2~mm continuum ring structure consists
of regularly spaced clumps which
implies the ``collect and collapse'' model of triggered star formation
(Elmegreen \& Lada 1977; Witworth et al. 1994; Dale et al. 2007).
According to this model the H~II region is expanding
into dense cloud and neutral material is accumulating
in the shell between shock and ionization fronts.
The shell splits into massive clumps by large-scale instabilities
which develop along the shell.
The clumps can be the sites of formation of the next generation of stars.
Low-mass clumps can be also formed by small-scale gravitational instabilities
(Deharveng et al. 2010).
The shells consisting of massive clumps
have been found on the periphery of several Galactic H~II regions
(e.g. Deharveng et al. 2003, 2008).
The masses of dust clumps in the W40 ring are typical for low-mass cores.
Nevertheless, their location along the ring is similar
to what is expected in the ``collect and collapse'' mode.
The presence of Class~I and Class~0 sources associated
with the ring clumps (Maury et al. 2011, Mallick et al. 2013)
indicate that star formation is already taking place there.

The comparison of the morphologies of ionized gas and dust
(Fig.~\ref{ionized_gas}, right panel)
indicates that the ring is probably not associated with
the main W40 H~II region as the driving source
(the massive star of the O9.5 spectral type associated
with IRS~1A~South according to Shuping et al. 2012,
Fig.~\ref{ionized_gas}, big star)
is located outside the area bounded by the ring.
A likely source for the ring formation
could be a distinct subregion of ionized gas associated with IRS~5
(the massive star of the B1V spectral type,
Shuping et al. 2012, Mallick et al. 2013,
Fig.~\ref{ionized_gas}, small star)
which lies to the north-west from IRS~1A~South.
The IRS~5 source is located within the area bounded by the ring
but away from the geometrical center of the ring
closer to the western branch (Fig.~\ref{ionized_gas}, right panel).
This could be a projection effect if the ring is not being viewed face-on.
Taking the maximum and minimum distances from IRS~5 to the inner boundary
of the ring $\sim 3'$ ($\sim 0.44$~pc) in the southern direction
and $\sim 46''$ ($\sim 0.11$~pc) to the west (Fig.~\ref{dust_sources})
an inclination angle of the ring is $\sim 75\degr$ in the azimuthal
plane assuming it is a circular in shape.
The south-eastern part of the ring is probably closer to the observer.

\subsection{The eastern branch of the ring}
\label{east_branch}

Close to the eastern branch of the dust ring
the VLA 3.6~cm and NIR sources are located.
The area of the dust clumps~2 and 3
has an elongated form in the north-south direction.
The maps of ionized gas and dense molecular gas traced by CS(5--4)
(Fig.~\ref{ionized_gas}) show similar shapes towards this area.
The similarities imply that they belong to the same region
where interaction between ionized and neutral material is taking place.
There is an enhanced 2.12~$\mu$m H$_2$ emission region
of a similar elongated form close to this part of the ring
(Mallick et al. 2013) indicating an existence of shocked gas.
We detected the narrow CH$_3$OH(2--1) line at $\sim 8$~km~s$^{-1}$
towards one of the eastern positions confirming an existence
of shocked gas or inner YSO.
The dust temperatures and molecular line widths are higher
in this area compared with the rest of the cloud.
The CS(2--1) line profiles are asymmetric implying contraction
which is confirmed by model simulations (Section~\ref{modeling}).
The power-law indices of systematic velocity and density radial profiles
in infalling gas derived from model simulations
differ from those of the ``inside-out'' model of low-mass star formation,
yet, the estimate of the mass falling rate is typical
for low-mass star formation regions.
The model simulations also imply an existence
of an extended low-density gas in this region.
Probably an ionized front from the main H~II region
or from the smaller region associated with VLA~3 expands
into the cloud towards the west where it is confronted with the dust ring.
A shock wave is formed at the outer boundary of the eastern branch
of the ring and compressed the material of the eastern clumps
and probably forcing contraction.

\subsection{Kinematics of the region}

The gas kinematics is rather complex.
Molecular lines at $\sim 4.5-5$~km~s$^{-1}$ are observed over
the region.
Towards the eastern branch of the dust ring
the CS and some other lines that can be associated with
contracting gas are observed at $\sim 7-8$~km~s$^{-1}$.
The CO emission at higher velocities is observed to the east
of the ring including the position of the main driving source.
Strong and wide OH absorption line at $\sim 7$~km~s$^{-1}$
observed towards the W40 H~II region by Crutcher (1977) implies that
the gas at this velocity is located in front of the H~II region.
Vall\'ee (1987) and Shuping et al. (1999) made a conclusion that
there should be amount of gas located in front of the H~II region.
This foreground gas could give emission in molecular lines
at $\sim 7$~km~s$^{-1}$ towards the eastern branch of the dust ring.
The gas of the main cloud with $\sim 4.5-5$~km~s$^{-1}$ velocities
is probably located at the background.
Yet, the relative line-of-sight location
of the gas at $\sim 7-8$~km~s$^{-1}$ and $\sim 4.5-5$~km~s$^{-1}$
is unclear towards the eastern branch of the ring
where two distinct CS(5--4) clumps with different
velocities are associated with the dust ring.

\subsection{Chemical composition}
\label{chemistry}

Molecular line observations revealed strong chemical differentiation
over the cloud.
The CS abundance in the contracting gas is $\sim 7-10$ times higher
than in ``quiescent'' gas towards western positions
and towards the dust clump~1.
The CS enhancement could be connected with evaporation of grain mantles
as predicted by hot core chemical models (e.g. Nomura \& Millar 2004).
An enhanced value of $X$(CS) is also derived towards the ``southern CS clump''.
This clump is not related to the ring having
lower molecular hydrogen column density compared with the ring clumps.
An absence of the HCN and HCO$^+$ emission at the velocities
of the infalling gas ($\sim 7-8$~km~s$^{-1}$) towards the eastern area
and detection of their rare isotopic lines at these velocities
imply some specific excitation conditions.
The reasons of this phenomenon could be connected with
absorption of optically thick
lines in extended low-density envelope and/or
with an influence of continuum emission and/or conditions
in the shocked gas on excitation of these molecules.
Yet, the spectra are not of the good quality to perform detailed analysis.

The emission from nitrogen-bearing molecules (NH$_3$ and N$_2$H$^+$)
is concentrated in clumps in the western branch of the dust ring.
Towards the dust clump~6
the N$_2$H$^+$, NH$_3$ and H$^{13}$CO$^+$ lines are more narrow
compared with the CS, HCN and HCO$^+$ lines.
The latter do not trace clumps
and probably arise in the surrounding gas with higher degree of turbulence.
In the other parts of the region the NH$_3$ and N$_2$H$^+$ lines
are weak and show no correlation with the ring structure.
The NH$_3$ and N$_2$H$^+$ abundances towards the western positions
are $\sim 3-10$ times higher than towards the other ones
(Table~\ref{abundances}).
The N$_2$H$^+$ abundance drop observed towards embedded massive YSOs
(e.g. Pirogov et al. 2007, Zinchenko et al. 2009, Reiter et al. 2011)
can be explained by N$_2$H$^+$ dissociative recombination.
Low NH$_3$ and N$_2$H$^+$ abundances
in the central and eastern parts of the W40 region
can be associated with high level of the ionized emission
from the cluster stars and IRS~5 which penetrates through
inhomogeneous surrounding gas enhancing electron abundances
and destroying nitrogen-bearing molecules
or reducing their production rates.

\subsection{Evolutionary status of the region}

The eastern branch of the ring
obviously differs in conditions from the other parts of the cloud
which could be connected with different evolutionary stages.
Class I sources are located near the eastern dust clumps~2 and 3
while Class 0 sources are close to the dust clumps~6 and 9
of the western branch (Maury et al. 2011).
The NH$_3$ and N$_2$H$^+$ emission
is concentrated mainly in the western branch,
while the CS emission is high in the eastern part of the cloud.
The eastern clumps~2 and 3 are more massive with higher extent
of non-thermal motions and higher dust temperatures.
The CS line profiles towards them possess
blue asymmetry.
It is known that blue excess is more common for the UC~HII regions
than for their precursors
(e.g. Wu et al. 2007, Fuller et al. 2005, Liu et al. 2011).
We suggest that the eastern branch is more evolved than the western one
and an evolution sequence is going from the east to the west,
from the eastern clumps more affected by the main driving source
to the remote western ones.
Such evolution sequence is seen in the surrounding of other massive stars
(Liu et al. 2012).
We suggest that the area of the eastern clumps~2 and 3
could probably be a site of triggered formation of the stars with masses
higher than one solar mass.

\vspace{3mm}

The questions which remain are
connected with detailed
morphology of the gas with different velocities especially
towards the eastern branch of the dust ring.
New multiline observations with high spatial resolution and sensitivity
are needed to study in detail the structure of the region
where interaction between ionized and neutral material is taking place.
The modeling of chemical and physical parameters
is needed to explain reduced abundances of nitrogen-bearing molecules
in the regions of enhanced ionization
and anomalies in the HCN/H$^{13}$CN and HCO$^+$/H$^{13}$CO$^+$
intensity ratios in the gas with enhanced dynamical activity.

\section{Conclusions}
\label{conclusions}

In order to study structure and physical properties of dense molecular gas
and dust associated with W40, one of the nearby H~II regions,
millimeter-wave molecular line and 1.2~mm continuum single dish
observations have been performed.
This region has been also observed
at the GMRT interferometer at 1280~MHz and 610~MHz.
The cloud is located to the north-west of the main H~II region
and has complex morphological and kinematical structure including
clumpy dust ring and extended dense core.
The ring is probably formed by the ``collect and collapse'' process
due to expansion of the smaller neighboring H~II region
associated with IRS~5 and not being viewed face-on.
The eastern branch of the ring is probably closer to us.
The results of the study can be summarized as follows:

1. Nine dust clumps in the ring have been deconvolved.
Their sizes, masses and peak hydrogen column densities
derived from the continuum data lie in the ranges: $\sim 0.02-0.11$~pc,
$\sim 0.4-8.1~M_{\odot}$ and $\sim (2.5-11)\times 10^{22}$~cm$^{-2}$,
respectively.
The most strong continuum emission comes from the eastern clumps 2 and 3.

2. Sharp differences in morphologies of molecular and dust maps are found.
Molecular lines at $\sim 4.5-5$~km~s$^{-1}$ are observed
over the whole region including the area without prominent dust emission.
Towards the eastern branch of the ring
the CS and some other lines are observed at $\sim 7-8$~km~s$^{-1}$.
The eastern dust clumps~2 and 3 are associated with the CS(5--4) clumps.
The NH$_3$, N$_2$H$^+$, and H$^{13}$CO$^+$ emission
is enhanced towards the western dust clumps.
The HCN(1--0) and HCO$^+$(1--0) maps show no correlation with dust.
Probably these lines and the CO lines (Zhu et al. 2006)
trace the surrounding gas.

3. The physical parameters are derived from molecular line data
towards selected positions with spatial extent of 40$''$
using LTE and non-LTE analysis.
Ammonia kinetic temperatures are 21~K and 16~K
for two western clumps.
Their virial masses,
$\sim 3~M_{\odot}$ and $\sim 2~M_{\odot}$, respectively,
are close to the masses derived from the dust continuum data.
Number densities derived from the CS modeling
lie in the range: $\sim (0.3-3.2)\times 10^6$~cm$^{-3}$.
The CS(2--1) line profiles
towards the eastern positions are self-reversed and asymmetric
implying infall motions.
The calculations within inhomogeneous
non-LTE spherically-symmetric model with
systematic velocity profile $\propto r^{-0.1}$
the density profile $\propto r^{-2}$,
and extended homogeneous low-density envelope
give good fits to the CS and C$^{34}$S profiles observed
towards one selected position.

4. Molecular column densities and abundances imply
strong chemical differentiation over the region.
The CS abundances are enhanced towards the eastern dust clump~2
and the southern CS clump.
The NH$_3$, N$_2$H$^+$, and H$^{13}$CO$^+$  abundances
towards the western positions
are up to an order of magnitude higher
than towards the eastern positions.
The electron abundances for two positions
are $\sim (2-3)\times 10^{-7}$.
Probably high $X$(e) values and low abundances
of nitrogen-bearing molecules
are related to inhomogeneous gas that surrounds the clumps.

5.
The area including the eastern clumps 2 and 3 differs
from the western one.
These clumps associated with Class~I sources
are more massive with higher extent of non-thermal motions
and higher dust temperatures.
An interaction between ionized and neutral material is taking place
in the vicinity of this area.
Dense molecular gas is probably contracting.
We suggest that the eastern branch of the ring is more evolved
than the western one and could probably be a site of triggered formation
of the stars with masses higher than one solar mass.

\section*{Acknowledgments}

We would like to thank Lei Zhu for providing
the CO observational data.
We are grateful to Guillermo Quintana-Lacaci for pre-processing
the MAMBO data and to Alex Kraus for calculation
atmospheric optical depths and system temperatures
for the Effelsberg observations.
We are also grateful to Alexander Lapinov for calculating
statistical weights of the hyperfine components
of the NH$_3$(1,1) and (2,2) transitions.
We would like to thank the anonymous referee
for the comments and recommendations that improved the paper.
The research made use of the SIMBAD database,
operated by the CDS, Strasbourg, France.
The work was supported by the Russian Ministry of Education and Science
(grant 8421) and by the Russian Foundation for Basic Research
of the Russian Academy of Sciences
(grants 12-02-00861, 11-02-92690 and 13-02-92697).

\label{lastpage}

\end{document}